\documentclass[a4paper,11pt]{article}
\pdfoutput=1 
\usepackage{jheppub} 
\usepackage{amsfonts}
\usepackage{textcomp}
\usepackage{amssymb,amsmath}
\usepackage{hyperref}
\usepackage{comment}
\usepackage{graphicx}
\usepackage{multirow,booktabs,multirow}
\usepackage{braket}
\usepackage{bbold}
\usepackage{xcolor}
\usepackage[utf8]{inputenc}
\usepackage{cleveref}
\usepackage{eqparbox}
\usepackage{cancel}

\usepackage{mathtools}

\usepackage[T1]{fontenc} %

\title{Polarimetric searches for axion dark matter and high-frequency gravitational waves using optical cavities}
\author[a]{Camilo Garc\'ia-Cely,}
\author[a]{Luca Marsili,}
\affiliation[a]{
Instituto de F\'isica Corpuscular (IFIC), Consejo Superior de Investigaciones
Cient\'ificas (CSIC) and Universitat de Val\`encia,  C/ Catedratico Jose Beltran 2, E-46980 Paterna, Spain}
\author[b]{Andreas Ringwald,}
\author[b]{Aaron D. Spector}
\affiliation[b]{Deutsches Elektronen-Synchrotron DESY, Notkestr. 85,
22607 Hamburg, Germany}

\abstract{
We revisit birefringence effects associated with the evolution of the polarization of light as it propagates through axion dark matter or the background of a passing gravitational wave (GW). We demonstrate that this can be described by a unified formalism, highlighting a synergy between searches for axions and high-frequency GWs. We show that by exploiting this framework, the optical cavities used by the ALPS II experiment can potentially probe axion masses in the range 
$m_a \sim 10^{-9} - 10^{-6} \, \mathrm{eV}$, offering competitive sensitivity with existing laboratory and astrophysical searches.  
Also building on this approach, we propose using these optical cavities to search for high-frequency GWs by measuring changes in the polarization of their laser.  This makes it a promising method for exploring, in the near future, GWs with frequencies above $100$ MHz and  strain sensitivities on the order of $10^{-14} \, \mathrm{Hz}^{-1/2}$. Such sensitivity allows the exploration of currently unconstrained parameter space, complementing other high-frequency GW experiments.
This work contributes to the growing community investigating novel approaches for high-frequency GW detection.

}


\begin{document}
\preprint{DESY-25-002}
\maketitle

\section{Introduction}

Axions are not only a compelling solution to the strong CP problem of the Standard Model~\cite{Peccei:1977hh, Vafa:1984xg} but also a well-motivated candidate for the dark matter (DM) of the Universe~\cite{Weinberg:1977ma,Wilczek:1977pj}. 
They can be searched for through their coupling to electromagnetic fields~\cite{Sikivie:1983ip}. For instance, electromagnetic waves propagating through a strong magnetic field can produce axions, which, unlike light, can pass unhindered through physical obstacles and be detected if they are converted back to electromagnetic waves. This leads to the `light-shining-through-a-wall' approach of axion searches~\cite{Anselm:1985obz,VanBibber:1987rq}, employed by experiments such as ALPS II~\cite{Bahre:2013ywa}.  For a review, see Refs.~\cite{Redondo:2010dp, ParticleDataGroup:2024cfk}.

Additionally, due to the same coupling to photons, an axion background acts as a circularly birefringent medium, causing light propagating through it to experience different phase velocities for each circular polarization~\cite{Maiani:1986md}. This has inspired numerous experimental proposals to search for axion DM using optical cavities, where this birefringent effect can be precisely measured~\cite{Cameron:1993mr, Tam:2011kw, DellaValle:2015xxa, Liu:2018icu,Martynov:2019azm, Obata:2018vvr, DeRocco:2018jwe, Nagano:2019rbw,  Nagano:2021kwx}.
Remarkably, this has also sparked a synergy between the communities searching for gravitational waves (GWs) and axions, as GW detectors such as LIGO and Virgo use powerful optical cavities that can also be repurposed for axion searches~\cite{DeRocco:2018jwe,Nagano:2019rbw,  Nagano:2021kwx, Gue:2024txz}.  

In this work, we explore an additional aspect of this synergy by highlighting that optical cavities optimized for detecting axion-induced birefringence can also function as probes for GWs. This is because a varying gravitational background can act as a birefringent medium, altering the polarization of light as it propagates through it. We refer the reader to  Ref.~\cite{Weinberg:2008zzc} for a textbook treatment. In fact, experiments leveraging this effect in resonant cavities have been proposed to detect GWs~\cite{ Iacopini:1979ww, Pegoraro:1978gv,Cruise1, Cruise2, Cruise3, Cruise:2006zt}. In this context, we present a unified formalism that describes the birefringent effects of both axion and GWs within the same framework.  Furthermore, we illustrate this for several optical cavities similar to those used in the ALPS II experiment, which could be adapted to measure polarization effects. In this way, we derive future sensitivity projections for detecting axion DM or high-frequency GWs and show that this approach is competitive with other experimental methods.
 
A compelling motivation for this investigation is the fact that no known astrophysical object is both small and dense enough to emit GWs at frequencies above 10 kHz. Detecting GWs at such high frequencies would therefore suggest the presence of physics beyond the Standard Model of particle physics. Indeed, recent years have seen a growing community~\cite{Aggarwal:2020olq} interested in searching for high-frequency GWs, driven by the goal of detecting early-universe signals predicted to exist by several models of physics beyond the Standard Model. Our work therefore complements efforts in this direction.

This work is organized as follows: In Section \ref{sec:basics}, we outline the fundamental principles of geometric optics that enable the study of the polarization evolution of light in a slowly varying background of axions or gravitational fields, with particular emphasis on placing both phenomena on a unified framework. In Section \ref{sec:application}, we apply these principles to resonant cavities, while in Section \ref{sec:ALPSII}, we focus on modifications to the ALPS II cavities and present the corresponding sensitivity prospects on the axion-photon coupling and GW strain. Finally, in Section \ref{sec:conclusions}, we provide our conclusions, and in the appendices \ref{app:A} and \ref{app:B}, we detail the geometric-optics limit in the presence of axions and GWs. Throughout we will adopt a Minkowski metric with $\eta_{\mu\nu} = \text{diag}(-+++)$ and work with natural Heaviside-Lorentz units ($\hbar=c=1$). Moreover, Greek indices always represent space-time variables, while Latin indices exclusively represent three-dimensional variables.

\section{Geometric optics in the presence of axion DM or GWs }
\label{sec:basics}
We aim to study the polarization of light as it propagates in the background of axion DM or a passing GW. We will focus on light with typical wavelengths, $\lambda = 1/f_L$, much smaller than the characteristic length scale, $d$, over which the background varies.
Specifically, in the context of an optical laser, which we will analyze below, the wavelength lies in the optical range, while $d$ is assumed to be in the millimeter range or larger. Then, the corresponding electromagnetic field  has a phase that changes very fast while  its amplitude remains nearly constant.  Under these circumstances, Maxwell's equations can be solved by formally casting the field as
\begin{align}
F^{\mu\nu} = (f^{\mu\nu}+{\cal O}(\epsilon)
+\ldots) e^{i \theta/\epsilon } \,, 
\label{eq:eikonal}
\end{align}
and expanding on $\epsilon$, a fictitious parameter eventually set to unity. This expansion is useful because  a term multiplied by $\epsilon^n$ is of order $(\lambda/d)^n$. The leading term in this expansion is the geometric-optics limit of electromagnetism~\cite{Maggiore:2007ulw}. For illustration, let us first examine geometric optics in flat spacetime, using a standard experimental setup as the background. While this analysis will simply show that the polarization remains constant as light follows null geodesics, it serves as a basis for generalizing to the cases of axions and GWs.
First, note that $\partial^\lambda F^{\mu\nu} =  i \left( f^{\mu\nu} \partial^\lambda \theta \right)\frac{1}{\epsilon}+{\cal O}(\epsilon^0)$. Defining $k_\nu  \equiv \partial_\nu \theta$, the ordinary  Maxwell's equations
\begin{align}
\label{eq:MaxwellFLAT}
\partial_\nu F^{\mu\nu} =0\,, && \partial^\lambda F^{\mu\nu}+\partial^\mu F^{\nu\lambda}+\partial^\nu F^{\lambda\mu}=0  \,,
\end{align}
lead to
\begin{align}
k_\nu f^{\mu\nu} = 0 \,, && \text{and}&&
k^\lambda f^{\mu\nu}+k^\mu f^{\nu\lambda}+k^\nu f^{\lambda\mu}=0\,.
\label{eq:appgeometricFLAT1}
\end{align}
Multiplying the second equation by $k_\nu$ and using the first relation, we obtain 
\begin{equation}
 k_\nu k^\nu=0   \,.
 \label{eq:k2}
\end{equation}
Hence, $k^\mu$ can be thought of as a four-momentum, defining the tangent vector of null geodesics
\begin{align}
k^\mu = \frac{dx^\mu}{d\ell} \,, && \text{or equivalently}&& \frac{dx^\mu}{dt} = \frac{k^\mu}{k^0}\,.
\label{eq:nullgeodesics}
\end{align}
On the other hand, the wave equation for the electromagnetic field --which is of course a consequence of Eq.~\eqref{eq:MaxwellFLAT}-- gives rise to 
\begin{eqnarray}
\partial_\rho \partial^\rho F^{\mu\nu} \,\nonumber= 
\partial_\rho \theta \partial^\rho \theta\left(- f^{\mu\nu} \frac{1}{\epsilon^2} +\cdots\right) +\left(2i \partial^\rho f^{\mu\nu} \partial_\rho \theta + i f^{\mu\nu} \partial_{\rho}\partial^{\rho}\theta  \right) \frac{1}{\epsilon} +{\cal O}(\epsilon^0) =0\,.
\end{eqnarray}
The first term always vanishes due to Eq.~\eqref{eq:k2},  while the second vanishes  only if 
\begin{equation}
k^\rho \partial_\rho f^{\mu\nu} = -\frac{1}{2} f^{\mu\nu} \partial^\rho k_\rho\,.
\label{eq:appgeometricFLAT2}
\end{equation}
In particular, if we define the polarization vector as the direction of the electric field 
\begin{align}
\label{eq:poldef}
 e^{\mu} \equiv\frac{f^{0\mu}}{\sqrt{ f^{0\nu}f^*_{0\nu}}} \,, 
\end{align}
Eq.~\eqref{eq:appgeometricFLAT2} indicates that  
$k^\rho \partial_\rho e^i= 0$. This can also be written as
\begin{eqnarray}
 \frac{de^i}{dt} =0\,,
\label{eq:final0}
\end{eqnarray}
because, according to Eq.~\eqref{eq:nullgeodesics}, along null-geodesics $ k^\rho \partial_\rho = k^0\frac{d}{dt}$. Although this analysis is purely classical, the interpretation of these equations—and consequently of geometric optics— can be phrased as follows: The vector \(k^\mu\) represents the four-momentum of photons, while \(e^i\) specifies their polarization, which remains unchanged as they propagate along null geodesics.

\paragraph{\bf The case of axion DM.}

The geometric-optics limit can also be applied to axion electrodynamics, that is, to 
the case where the Lagrangian of Maxwell's equations is augmented with
\begin{equation}
\label{eq:Maxwell}
{\cal L}=-\frac{1}{4} g_{a \gamma \gamma} a(t) F^{\mu \nu} \tilde{F}_{\mu \nu}=g_{a\gamma \gamma} a(t) {\bf E}\cdot{\bf B} \,,
\end{equation}
where $a(t)$ is a slowly-changing axion field. If axions comprise the DM of the galaxy
\begin{equation}
a(t) =a_0 \sin(m_a t+ \varphi ) \,,
\label{Eq:Ax_A}
\end{equation}
where $m_a$ is the axion mass and $a_0$  is fixed by the local DM density according to $\rho_a = a_0^2 m_a^2/2 = 0.3~\mathrm{GeV/cm^3}$ (see, e.g.,~\cite{ParticleDataGroup:2024cfk}).
Nevertheless,  due to their velocity dispersion in the galaxy, $v_a$, the axion field is coherent only for a time~\cite{ParticleDataGroup:2024cfk}
\begin{equation}
\tau  = \frac{2\pi}{m_a v_a^2} \approx\frac{10^{-16} {\rm eV}}{m_a} \, {\rm year}\,.
\label{eq:axiontau}
\end{equation}

To compute the evolution of the polarization vector in the geometric-optics limit, we follow the same procedure outlined above, with further details provided in Appendix~\ref{app:A}.  In particular, we find that Eq.~\eqref{eq:final0} generalizes to

\begin{align}
\frac{d{\bf e}}{dt} = 
 - \frac{1}{2}g_{a\gamma\gamma}  \dot{a}(t) \, \hat{\bf k} \times {\bf e} \,.
 \label{eq:axionsGEOMETRIC}
\end{align}
If the initial polarization is linear, this equation states that the vector ${\bf e}$ rotates with an angular velocity $g_{a\gamma\gamma}  \dot{a}(t)/2$ around the direction $\hat{\bf k}$, while its absolute magnitude and the angle with respect to this direction do not change.  For right (left) circular polarizations
$
\hat{\bf k} \times {\bf e} = -i \lambda \, {\bf e}
$
where $\lambda$ is $+1$ ($-1$). At leading order in $g_{a\gamma\gamma}$, Eq.~\eqref{eq:axionsGEOMETRIC} 
implies that left- and right-polarized light propagates with different phase velocities, namely $1+\lambda \delta c (t)$, with  
\begin{equation}
\delta c(t) = \frac{g_{a\gamma\gamma}  \dot{a}(t)}{2\omega_L} 
= \frac{g_{a\gamma} \rho_a}{\sqrt{2} \omega_L} \cos(m_a t+\varphi)\,,
\label{eq:defc}
\end{equation} 
where $\omega_L =2\pi f_L$ is the frequency of the electromagnetic wave.  
This is the origin of the term birefringence, which resembles the Faraday effect, a magneto-optical phenomenon where the polarization of linearly polarized light rotates as it propagates through a magnetic field. In fact, such an effect arises from an equation completely analogous to Eq.~\eqref{eq:axionsGEOMETRIC}. 
\paragraph{\bf The case of curved spacetimes.} The geometric-optics formalism leading to Eqs.~\eqref{eq:appgeometricFLAT1} and \eqref{eq:appgeometricFLAT2} can also be extended to a curved spacetime, and in particular, to GWs. 
For this, we note that the corresponding Maxwell's equations can be obtained from those in flat spacetime replacing ordinary partial derivatives by covariant derivatives.  Hence, in an arbitrary spacetime the equation of geometric optics read\footnote{In this regard, it is important to mention that the derivation of Eqs.~\eqref{eq:appgeometricFLAT1} and \eqref{eq:appgeometricFLAT2} was done without assuming $\partial^\alpha \partial^\beta =\partial^\beta \partial^\alpha$ and are therefore equally valid for covariant derivatives.}
\begin{align}
k_\nu f^{\mu\nu} = 0 \,, && 
k^\lambda f^{\mu\nu}+k^\mu f^{\nu\lambda}+k^\nu f^{\lambda\mu}=0\,,
&&\text{and}&&
k^\rho \nabla_\rho f^{\mu\nu} = -\frac{1}{2} f^{\mu\nu} \nabla^\rho k_\rho\,.
\label{eq:appgeometric3}
\end{align}
As before, they also imply 
\begin{equation}
k_\mu k^\mu=0\,. 
\label{eq:null-geodesics}
\end{equation}
Moreover, a careful algebraic manipulation of Eq.~\eqref{eq:appgeometric3}, outlined in the Appendix~\ref{app:B},  shows that 
\begin{eqnarray}
k^\rho \left(\partial_\rho e^{i} -\frac{1}{k^0} \Gamma^{0}_{\rho \lambda}k^i e^{\lambda}+\Gamma^{i}_{\rho \lambda}e^{\lambda}\right) = 0\,,
\label{eq:GWsGEOMETRIC}
\end{eqnarray}
where $e^\mu$ is defined as in Eq.~\eqref{eq:poldef}.  In particular, $e^0=0$ everywhere. We remind the reader that the position of indices is now important.
 As above, along null-geodesics,  Eqs.~\eqref{eq:nullgeodesics} and  \eqref{eq:GWsGEOMETRIC} give the evolution of the polarization vector as%
\begin{eqnarray}
 \frac{de^i}{dt} = \left(\Gamma^{0}_{\rho \lambda} \frac{dx^i}{dt} -\Gamma^{i}_{\rho \lambda} \right)\frac{dx^\rho}{dt}e^{\lambda} \,. %
\label{eq:finalGW}
\end{eqnarray}
This must be compared against the analogous Eqs.~\eqref{eq:axionsGEOMETRIC} and \eqref{eq:final0} for a flat spacetime with and without axions, respectively.  As usual, the metric perturbation, $h_{\mu\nu}$, associated with the GW enters in the Christoffel symbols as
\begin{equation}
\Gamma^\mu_{ \rho \lambda} = \frac{1}{2} \eta^{\mu \sigma} \left( \partial_\rho h_{\sigma \lambda} + \partial_\lambda h_{\sigma \rho} - \partial_\sigma h_{\rho \lambda } \right).
\end{equation}
We are interested in applying Eq.~\eqref{eq:finalGW} to a passing GW in the transverse-traceless (TT) frame. Assuming it comes from a fixed direction, $\hat{\bf q}$, and allowing for an arbitrary frequency
\begin{align}
    h^{ij}(t,{\bf x}) = \int_{-\infty}^{\infty} df\,\tilde{h}(f)   e^{-2i\pi f  (t - \hat{\bf q} \cdot {\bf x})} e^{ij}(\hat{\bf q}) \,,%
    \label{eq:homega}
\end{align}
in which
 \begin{align}
e^{ij}(\hat{\bf q})&\!\!=  
\left\{
\begin{array}{ll} 
\frac{{\rm U}_i {\rm U}_j\!-\!{\rm V}_i {\rm V}_j}{\sqrt2}\, \quad\text{($+$ polarization for the GW)}, \\ 
\frac{{\rm U}_i {\rm V}_j\!+\!{\rm V}_i {\rm U}_j}{\sqrt2}\, \quad\text{($\times$ polarization for the GW)},  
\end{array}
\right.
\label{eq:planewave}
\end{align}
with
 \begin{align}
\hat{\bf q}\!\!= \sin\theta_h\,\hat{\mathbf{e}}_\rho +\cos\theta_h\,\hat{\bf e}_z,\,\,
&&
\mathbf{V} = \hat{\mathbf{e}}_{\phi_h},\,\,
&&
\mathbf{U} = \mathbf{V} \times \hat{\bf q}.
\end{align}
Finally, let us note that, for a passing GW, not only does the polarization vector change, but also the corresponding momentum, $k^\mu$. This  can be seen easily  from $\left(\eta_{\mu\nu} +h_{\mu\nu}\right) k^\mu k^\nu=0$, as follows from Eq.~\eqref{eq:null-geodesics}. The solution to this equation is known to be given by the parallel transport of the four momentum along null-geodesics
\begin{align}
\label{eq:finalGWk}
\frac{d k^\mu}{dt}  = - \Gamma^\mu_{\rho\lambda} \frac{dx^\rho}{dt} \frac{k^\lambda}{k^0} \,.
\end{align}
This should be compared with Eq.~\eqref{eq:finalGW}, which has an extra term whose origin has to do with the fact that $e^0=0$. More specifically, it stems from the fact that $ e^\mu = f^{0\mu}$ is not a true vector.\footnote{For a textbook discussion in the context of the CMB polarization, see Appendix G of Ref.~\cite{Weinberg:2008zzc}, which  derives Eq.~\eqref{eq:finalGW} exploiting this property. 
}     
Moreover, it is possible to show from Eqs.~\eqref{eq:finalGW} and \eqref{eq:finalGWk} that 
\begin{align}
\label{eq:kidentitites}
k_i e^i=k_\mu e^\mu=0 \,, && \left(\delta_{ij}+ h_{i j}\right) e^i e^{j *}= \left(\eta_{\mu\nu}+ h_{\mu \nu} \right)e^\mu e^{\nu *}=1 \,.
\end{align}
Hence, as expected for polarization vectors, their evolution ensures that they remain orthogonal to the momentum, that also changes due to the parallel transport in Eq.~\eqref{eq:finalGWk}. 
\paragraph{\bf Unified treatment.}
The evolution of the polarization vector in the presence of a background of axion DM or GWs can then be expressed as
\begin{align}
\frac{d{e^i}}{dt} = {\mathcal{M}}^{ij}(t)  \, e^j\,,&&\text{with}
&&  M^{ij}(t)\equiv\int_{-\infty}^{\infty}df \,  e^{- 2i\pi f t}   \tilde{\cal M}^{ij} \,.
 \label{eq:M}
\end{align}
A detailed calculation based on  Eqs.~\eqref{eq:axionsGEOMETRIC} and \eqref{eq:finalGW} yields 
 \begin{align}
\!\!\!\!\!\tilde{\cal M}^{ij} = %
\left\{
\begin{array}{ll} 
 - \delta\tilde{c}(f) \epsilon^{ijn} k^n \, 
 &\quad\text{for axions} , \\ 
 &\\
  \left[   (1-\hat{\bf k}\cdot \hat{\bf q}) \,  {e}^{ij}(\hat{\bf q}) -{e}^{in}(\hat{\bf q}) \hat{k}^n \hat{q}^j +  {e}^{jn}(\hat{\bf q}) \hat{k}^n \hat{q}^i -  \, {e}^{jn }(\hat{\bf q}) \hat{k}^n  \hat{k}^i \right]\,&\quad\text{for GWs}.\\ 
\times  i \pi  f \tilde{h}(f) e^{ 2i\pi f  \,\hat{\bf q} \cdot {\bf x}(t)} & 
\end{array}
\right.
\label{eq:Mtilde}
\end{align}
Here $\delta \tilde{c}(f)$ is the ordinary Fourier transform of $\delta c(t)$, defined in terms of the axion field in Eq.~\eqref{eq:defc}. 

In contrast,  $\tilde{\cal M} \big|_{\rm GW}$ is not a proper Fourier transform because of its time-dependent term $e^{2i\pi f \,\hat{\bf q} \cdot {\bf x}(t)}$, where ${\bf x}(t)$ is the photon trajectory and ${\bf \hat{k}} = d {\bf x}/dt + {\cal O}(\tilde{h})$,  as defined in Eq.~\eqref{eq:nullgeodesics}.
We nevertheless keep this notation as it will be useful below.
The previous equations motivate to define the {\it signal} 
 \begin{align}
s (f) =  
\left\{
\begin{array}{ll} 
\delta \tilde{c} (f) \, &\quad\text{for axions}, \\ 
\tilde{h}(f)\,& \quad\text{for GWs}. 
\end{array}
\right.
\label{eq:epsilon}
\end{align}
For instance, for a monochromatic signal --and assuming it to be a cosine-- we have  
\begin{equation}
s(f)= \frac{1}{2} s_0    \delta(f-f_0)+ (f\to -f) \,,
\label{eq:epsilonmono}
\end{equation}
 where $f_0$ is either the GW frequency or $m_a/2\pi$, and $s_0$ represents the amplitude of either the GW or the axion-induced phase velocity difference (see Eq.~\eqref{eq:defc}).

\section{Application to resonant cavities}
\label{sec:application}

\begin{figure}[h!]
    \centering   \includegraphics[width=0.99\textwidth]{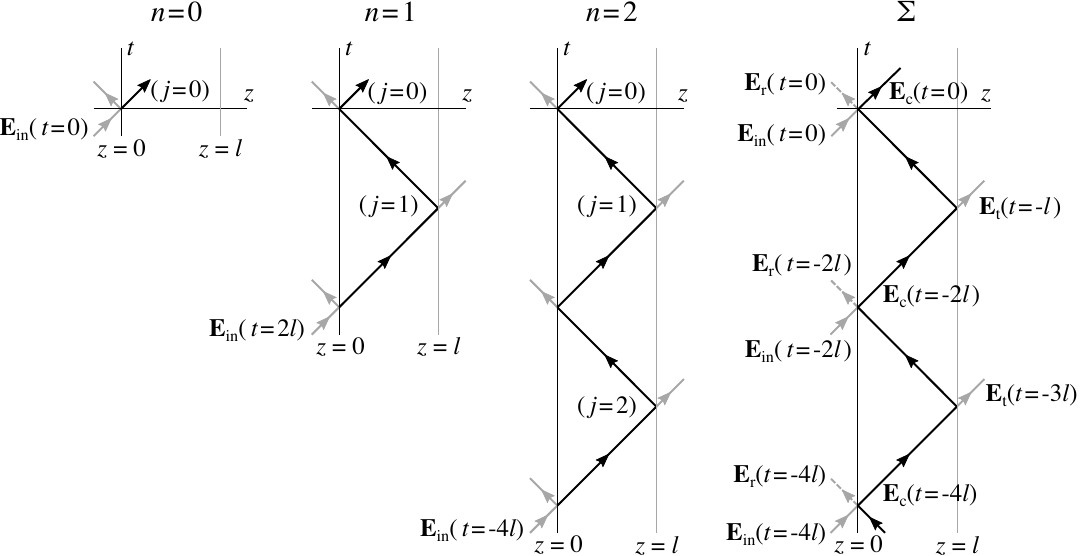}
    \caption{The laser beam within the cavity can be seen as a superposition of waves that enter the cavity at $t= -2nl$. Here $n$ labels the time at which the waves enter the cavity and $j$ labels each bounce. For example, the wave that enters at $t = -4l$ corresponds to the $n=2$ term in the sum of Eq.~\eqref{eq:eansatz} and travels two round trips labeled as $j=1$ and $j=2$. The full sum is then shown as the diagram on the right.}
    \label{fig:ang_plot}
\end{figure}

\subsection{Evolution in the absence of a background} 
Before discussing a more realistic setup in the next section, let us illustrate the use of Eq.~\eqref{eq:M} for a simplified configuration: a $p$-polarized laser approaching the region between two stationary mirrors, i.e. an optical cavity, propagating back and forth multiple times before exiting, see Fig.~\ref{fig:ang_plot}.  
With the notation introduced above the polarization vector associated with the laser is described\footnote{\label{footnote:f4}Throughout, boldface vectors are contravariant, that is $\bf e$ refers to $e^i$.} by ${\bf \hat{p}} = (1,0,0)$. If no axion DM  or a GW is present, the resulting light remains $p$-polarized.
To see this more clearly and to introduce notation that will be useful later, let us calculate  the resulting electric field  at mirror 1. We adopt a coordinate system such that, in the absence of axions or GWs, the laser  propagates along the $z$-direction starting at $t=-\infty$, with the two mirrors located at $z=0$ and $z=l$. As the field enters the cavity, it is attenuated by the transmission coefficient of the first mirror. Once the field is inside the cavity, it then propagates the distance $l$ between the mirrors and is reflected by the second mirror, receiving a small attenuation by the reflection coefficient of the mirror. The light then propagates back to the first mirror, where it is reflected and again attenuated slightly. Thus with each round trip through the cavity, a small portion of power leaves the field, either through transmission at one of the mirrors, or via some unintended loss channel such as scattering at the mirror surface, absorption in the mirror coatings, or clipping in beam tube. The field transmission coefficients for the mirrors will be referred to as $t_1$ and $t_2$, such that they have power transmissivity of $t_1^2$ and $t_2^2$, respectively. The total optical losses in field through one full round trip through the cavity will be denoted by $\ell$, with $\ell^2$ giving the total excess power lost per round trip. We will define the reflectivity coefficients of the mirrors using the following equations such that they include the round trip excess optical losses
\begin{align}
    1  = r_i^2 + t_i^2 + \ell^2/2\,,&&\text{with}&&i =1,2\,.
\end{align}
After completing a round trip through the cavity, the field interferes with the newly injected field at the first mirror. Under these circumstances, at $t=0$,  the electric field at mirror 1 is the superposition of several waves, each of which entered the region at $t=-2nl$ for certain integer $n$. Hence
\begin{align}
{\bf E}_1 (0)   
= \, \left( \sum^{\infty}_{n=0} e^{i\phi_n} \prod^{ n}_{j = 0} L_j  \right) \, t_1{\bf E}_{\rm in}(0)\,,
\label{eq:eansatz}
\end{align}
with
\begin{align}
\phi_n= 2n\omega_L l
\,, && \text{and} &&
L_j = r_1
    \mathcal{P} \times r_2 \mathcal{P} 
\,.
\label{eq:eansat_vacuum}
\end{align}
Here, ${\bf E}_{\rm in}(0)$ is the incoming field, see Fig.~\ref{fig:ang_plot}, while ${\bf E}\to L_j {\bf E}$ represents the change in the electric field after one complete round trip through the cavity.\footnote{For a cavity with no time-dependent birefringence, $L_j$ does not depend on the index $j$, however for a background axion field or GW, we will see that that is not the case.} In addition, $\phi_n$ is the relative phase  accumulated by the field  from $t=-2nl$ to $t=0$, while
${\cal P}$  is a matrix giving the specular image of the photon three-momentum.
This matrix inverts the three-momentum while leaving the vectors normal to it invariant. Due to this,
\begin{equation}
    {\cal P }^2 = \mathbb{I}\,.
    \label{eq:Pproperty}
\end{equation}
Putting these results together, we obtain the well-known result
\begin{align}
{\bf E}_1 (0) 
=\left( \sum^{\infty}_{n=0} (r_1 r_2 \,e^{2 i \omega_L l} )^{n} \mathbb{I}  \right) \,t_1 {\bf E}_{\rm in}(0) = \frac{t_1}{1-r_1 r_2 \,e^{4 i \pi f_L l \,} }  {\bf E}_{\rm in}(0) \,.
\label{eq:Enobkg}
\end{align}

Therefore, due to the reflection within the mirrors, the incoming electric field is amplified, with a magnitude that reaches a maximum value when $2 f_L l \in \mathbb{Z}$. If ${\bf E}_0 \propto \hat{\bf p}$ as we assume here, no $s$-polarized component is obtained as $ {\bf \hat{s}} ^\dagger \cdot {\bf E} =0$ (where we define ${\bf \hat{s}}  = (0,1,0)$).  The frequency spacing between these resonances is known as the free spectral range (FSR).

\subsection{The effect of axion DM  or  GWs} 
Eq.~\eqref{eq:eansatz} can be extended to account for the presence of axion DM or GWs. To calculate the corresponding $L_j$ and $\phi_n$, let us list the possible effects resulting from them:
\begin{itemize}
\item {\bf Motion of the mirrors:} Axions do not move the mirrors. For GWs, we are interested in frequencies high enough that the experimental apparatus (mirrors, lasers, etc.) can be considered in free fall. Specifically, this requires GW frequencies to be much higher than those of the mechanical resonances (see e.g.~\cite{Ratzinger:2024spd}), which imposes a lower bound of the order \( v_s / d \), where \( v_s \) is the sound speed and \( d \) represents the characteristic device size. Conservatively assuming \( d \) is on the order of a few centimeters and taking a typical \( v_s \sim 10^3 \, \mathrm{m/s} \), the free-fall limit sets a lower frequency of approximately \( 10^5 \, \mathrm{Hz} \). Moreover, as is well-known, in the TT frame adopted here,  objects in free fall initially at rest do not move\footnote{
We emphasize that this depends on the adopted coordinate frame. For instance, in the proper detector frame, the mirrors do move if the apparatus is in free fall.
} when a GW passes~\cite{Maggiore:2007ulw}. Therefore, we can assume that the mirrors do not move for axion DM or GWs with frequencies above  \( 10^5 \, \mathrm{Hz} \).
Hence, as above, at $t=0$ the light at mirror 1 is a superposition of all waves that enter the cavity at $t=-2nl$, see Fig.~\ref{fig:ang_plot}.

\item {\bf Polarization change:}  As light propagates between points $a$ and $b$, its electric field, ${\bf E}=|{\bf E}|{\bf e}$, changes due to variations in both magnitude and direction
\begin{equation}
 { \bf E }_b 
 = { \bf E }_a+  \int^b_a dt' \left(\frac{d { \bf |E| }}{dt} {\bf e}+  |{\bf E}|  {\cal M} {\bf e}\right)  \,,
 \label{eq:propagationEq}
\end{equation}
with the matrix ${\cal M}$ given by \eqref{eq:M}. While for axions this matrix does not depend on the photon trajectory, for GWs it does, requiring careful tracking the back-and-forth bounces of light within the mirrors. This marks a qualitatively important distinction between axions and GWs, that will be detailed below together with the relevant limits of integration and the argument of ${\cal M}$. Moreover, as we expect the axion to be weakly coupled and  the GW to have a small amplitude,  Eq.~\eqref{eq:propagationEq} can be simplified by expanding on the signal, $s(f)$, introduced in Eq.~\eqref{eq:epsilon}.
Then, since both ${\cal M}$ and $d|{\bf E}|/dt$ are proportional to $s(f)$, the polarization vector in the integrand of Eq.~\eqref{eq:propagationEq} can be replaced by its initial value at $a$. Hence
\begin{equation}
{ \bf E }_b ={ \bf E }_a+\left(\int^b_a dt' \frac{d { \bf |E| }}{dt}  \mathbb{I}+\int^b_a dt'  |{\bf E}|  {\cal M}\right) {\bf e}_a = \left(\left(\cdots\right) \mathbb{I} + \int^b_a dt'    {\cal M}\right) {\bf E}_a \,.
\end{equation}
As explained below, and similarly to the case without a background, the term proportional to $\mathbb{I}$  does not contribute to the overall change in polarization.  Consequently, its coefficient will not be specified hereafter.\footnote{This coefficient is essential for calculating the effect of the background on the $p$-polarization. Beyond the geometric limit, it provides an additional enhancement~\cite{Domcke:2024abs}.}

\item {\bf Reflection and transmission:}  For the case of GWs, as explained above, the experimental apparatus and its components are assumed to be in free fall. This allows the use of standard Fresnel's laws for reflection and transmission,  which can be phrased 
$ {\bf E}\to {\cal P} {\bf E}$, 
where the matrix $\cal P $ is associated with the specular image of the three-momentum of photons. Although this momentum is affected by the GW according to  Eq.~\eqref{eq:finalGWk}, the resulting matrix can be cast as
\begin{equation}
{\cal P} =\begin{pmatrix}
1 & 0 & 0 \\
0 & 1 & 0 \\
0 & 0 & -1
\end{pmatrix} + {\cal O}(s) \,.
\label{eq:reflop}
\end{equation}
As axions do not affect the reflection of the laser,  this equation also applies in their case (in fact, without the ${\cal O}(s)$ piece). 
We also note that at zeroth order in $s(f)$, the distinction between covariant and contravariant indices in $ {\bf E}\to {\cal P} {\bf E}$ is inconsequential.
\item {\bf Phase shift:} The matrix ${\cal M}$, by construction, does not account for the global phase in Eq.~\eqref{eq:eikonal} associated with the propagation of the beam entering the cavity at $t = -2nl$. To account for this, in Eq.~\eqref{eq:eansatz} we will include
\begin{equation}
\phi_n=  2n \omega_L l  + {\cal O}(s)\,.
\label{eq:phaseshift}
\end{equation}

We note that the standard technique for detecting GWs relies on measuring the ${\cal O}(s)$ contribution. However, after accounting for the polarization change, this becomes a second-order effect, as will be clarified below. Therefore, we do not specify it further.
\end{itemize}
Having listed all the effects associated with an axion or GW background, we can now generalize Eq.~\eqref{eq:eansatz}.
Discriminating its different contributions, the matrix $L_j$, which accounts for the change in the electric field after one bounce, is then given by

\begin{eqnarray}
    L_j &=& \underbrace{r_1 
    \mathcal{P} }_{\text{Reflection off mirror 1}} \times  \underbrace{\left( \left(\cdots\right)\mathbb{I} + \int^{l}_{0}  dt' \,  {\cal M} (t'-(2 j\!+\!1)l)\Bigg|_{{\bf x}(t') = \hat{{\bf z}} \,(l-t') } \right)}_{\text{ Path from mirror 2 to mirror 1
} }  \nonumber \\ &\times& \, \underbrace{r_2 \mathcal{P}
}_{\text{Reflection off mirror 2}} \,  \times \underbrace{\left(\left(\cdots\right)\mathbb{I} + \int^{l}_{0 }  dt'  {\cal M} (t' -  (2j\!+\!2) l)\Bigg|_{{\bf x}(t') = \hat{{\bf z}} \,t' } \right)}_{\text{ Path from mirror 1 to mirror 2}}\,.
\label{eq:master_form_L}
\end{eqnarray}
The arguments of ${\cal M}$ can be readily understood from Fig.~\ref{fig:ang_plot}. Keeping only linear terms on $s(f)$ and exploiting Eq.~\eqref{eq:Pproperty},  we find
\begin{eqnarray}
\!\!\!\!\!L_j\! = \left( \cdots\right)\mathbb{I}\!+  \!r_1 r_2 \left(\int^{l}_{0} \!\! dt' \,  \mathcal{P}{\cal M} (t'\!-\!(2 j+1)l) \mathcal{P}\Bigg|_{{\bf x}(t') = \hat{{\bf z}} \,(l-t') } \! \!\!\!\!\!+
  \int^{l}_{0 } \!\!\! dt'    {\cal M} (t' \!-\!  (2j+2) l)\Bigg|_{{\bf x}(t') = \hat{{\bf z}} \,t'}   \!\right). \nonumber
\end{eqnarray}
The product entering in Eq.~\eqref{eq:eansatz} can be performed in a similar manner
\begin{eqnarray}
\prod^{n-1}_{j = 0}  L_j  =\left(\cdots\right) \mathbb{I}+(r_1 r_2)^n \left[ \sum^{ n-1}_{j = 0} \left(\int^{l}_{0}  dt' \,  \mathcal{P}{\cal M} (t'-(2 j+1)l) \mathcal{P}\Bigg|_{{\bf x}(t') = \hat{{\bf z}} \,(l-t') }  \right. \right. \nonumber \\  
\left. \left.  \!\!\!\!\!\!\!\!\!\!\!\! +\int^{l}_{0 } dt'    {\cal M} (t' -  (2j+2) l)\Bigg|_{{\bf x}(t') = \hat{{\bf z}} \,t'} \right) \right]. 
\label{eq:product0}
\end{eqnarray}
Meanwhile, employing the matrix $\tilde{\cal M}$ introduced in Eqs.~\eqref{eq:M}, we find
\begin{eqnarray}
 {\cal M} (t' -  (2j\!+\!2) l)\Bigg|_{{\bf x}(t') = \hat{{\bf z}} \,t' } = \int^{\infty}_{-\infty}  df e^{-2i\pi  f (t'-(2j\!+\!2) l)}  \tilde{\cal M} \Bigg|_{{\bf x}(t') = \hat{{\bf z}} \,t' }\,,
\end{eqnarray}
\begin{eqnarray}
\mathcal{P}{\cal M} (t'-(2 j+1)l) \mathcal{P}\Bigg|_{{\bf x}(t') = \hat{{\bf z}} \,(l-t') } =  \int^{\infty}_{-\infty}  df e^{-2 i \pi f(t'-(2j\!+\!1) l)}  \mathcal{P} \tilde{\cal M} \mathcal{P}\Bigg|_{{\bf x}(t') = \hat{{\bf z}} \,(l-t') }  \,.
\end{eqnarray}
In this form the sum over $j$ of Eq.~\eqref{eq:product0} can be performed easily,  yielding the following result
\begin{eqnarray}
\prod^{ n-1}_{j = 0}  L_j =\left(\cdots\right)\mathbb{I}& -& (r_1 r_2)^n \Bigg[ \int_{-\infty}^{\infty} df \frac{e^{4 i \pi  f l n}-1}{e^{-4 i \pi  f l}-1}
 \\
&&
\times \int^{l}_{0} dt' e^{-2i\pi  f t'}\left( e^{-2\pi i f l  }  \,  \mathcal{P}\tilde{{\cal M}}  \mathcal{P} \bigg|_{{\bf x}(t') = \hat{{\bf z}} \,(l-t') }\!\!\!\! +        \tilde{{\cal M}} \bigg|_{{\bf x}(t') = \hat{{\bf z}} \,t'} \right)\Bigg]\,. \nonumber 
\end{eqnarray}
According to Eq.~\eqref{eq:eansatz}, the total field inside the cavity is given by
\begin{eqnarray}
 {\bf E}_1 (0)   
&=& \left(\cdots\right) t_1{\bf E}_{\rm in}+ t_1\sum^{\infty}_{n=1} e^{ i \phi_n \,} 
  (r_1 r_2)^n \! \Bigg[  {\bf E}_{\rm in}
 \\
&& - \int_{-\infty}^{\infty} df \frac{e^{4 i \pi f l n  }-1}{e^{-4 i \pi f l  }-1} \int^{l}_{0} dt' e^{-2i \pi f t'}\left( e^{-2i\pi f l  }  \,  \mathcal{P}\tilde{{\cal M}}  \mathcal{P} \bigg|_{{\bf x}(t') = \hat{{\bf z}} \,(l-t') }\!\!\!\! +        \tilde{{\cal M}} \bigg|_{{\bf x}(t') = \hat{{\bf z}} \,t'} \right)  {\bf E}_{\rm in}\Bigg]\,. \nonumber 
\label{eq:Efield0}
\end{eqnarray}
The corresponding $s$-polarized piece is therefore
\begin{eqnarray}
{\bf \hat{s}} ^\dagger \cdot {\bf E}_1 (0) &=&
- \sum^{\infty}_{n=1} (r_1 r_2)^n  e^{i \phi_n} \int_{-\infty}^{\infty} df \frac{e^{4 i \pi  f l n}-1}{e^{-4 i \pi  f l}-1}\\
&&
\times {\bf \hat{s}} ^\dagger   \left( e^{-2i\pi f l  }\int^{l}_{0}  dt' \, e^{-2i\pi f t'}  \,  \mathcal{P}\tilde{{\cal M}}  \mathcal{P} \Bigg|_{{\bf x}(t') = \hat{{\bf z}} \,(l-t') }\!\!\!+  \int^{l}_{0 } \!\! dt'  e^{-2i\pi f  t'}\,     \tilde{{\cal M}} \Bigg|_{{\bf x}(t') = \hat{{\bf z}} \,t'} \right)\,t_1{\bf E}_{\rm in}\,. \nonumber 
\end{eqnarray}
\begin{table}[t]
    \centering
    \begin{tabular}{c|c}
        & ${\cal H }(f)/{\cal H }_0(f)$\\
        \hline\hline
   \multirow{2}{*}{ Axions} & 
    \multirow{2}{*}{
        $-\frac{i f_L}{f} \left( 1- e^{-2i \pi f l  } \right)^2 $ } \\
        & \\\hline
    \multirow{2}{*}{$h_\times$}  &
    \multirow{2}{*}{
       $  \frac{  (1-e^{- 2i \pi f l} )^2+2 e^{- 2i \pi f l}(1 - e^{2i \pi f l \cos\theta_h} )+ 
   2 (1-e^{-4i \pi f l} ) \cos\theta_h \cos^2\phi_h }
   {2 \sqrt{2}}$}\\
   & 
   \\
   
    \multirow{2}{*}{$h_\times (f = 1/2l)$ }  &
    \multirow{2}{*}{
       $  \frac{ 1+ e^{i \pi  \cos\theta_h} }
   {\sqrt{2}}$}\\
   & 
   \\
   
    \multirow{2}{*}{$h_\times (f=1/l)$}  &
    \multirow{2}{*}{
        $\frac{ 1-e^{2 i \pi  \cos\theta_h} }
   {\sqrt{2}}$}\\
   & 
   \\\hline
  \multirow{2}{*}{  $h_+$} &
    \multirow{2}{*}{
        $  (   1-e^{- 4i \pi f l}) \,\frac{      \, (3 + \cos 2 \theta_h) \sin2 \phi_h}
{8 \sqrt{2}}$ }   \\
&\\
  \multirow{2}{*}{  $h_+ (f=1/2l)$ } &
    \multirow{2}{*}{
        $0$ }\\
   \\
  \multirow{2}{*}{  $h_+ (f=1/l)$} &
    \multirow{2}{*}{
       $0$} 
       \\
& \\\hline\hline
     \multirow{2}{*}{ Axions (QWP) } & 
    \multirow{2}{*}{
        $ -\frac{i f_L}{f} \left( 1- e^{-4i \pi f l  }  \right)$ }   \\        &\\\hline
 \multirow{2}{*}{ $h_\times$ (QWP)} &
    \multirow{2}{*}{
$
    \frac{1-e^{-4i \pi f l}+2 \left(1+e^{-4i \pi f l}-2 e^{-2i \pi f l (1-\cos\theta_h)}\right) \cos \theta_h \cos^2\phi_h}{2 \sqrt{2}}$
    }
\\
& 
   \\
   \multirow{2}{*}{  $h_\times$ (QWP, $ f=1/2l)$} &
    \multirow{2}{*}{
$
    \sqrt{2}\left(1+e^{i \pi  \cos\theta_h} \right)\cos \theta_h \cos^2\phi_h$
    }
\\
 & 
   \\
      \multirow{2}{*}{  $h_\times$ (QWP, $ f=1/l)$} &
    \multirow{2}{*}{
$
    \sqrt2 \left(1- e^{2i \pi  \cos\theta_h}\right) \cos \theta_h \cos^2\phi_h$
    }
\\
 & 
   \\\hline
 \multirow{2}{*}{$h_+$ (QWP)}   &
    \multirow{2}{*}{
$ \left(1 + e^{-4 i \pi f l} -2 \, e^{-2i\pi f l (1-\cos\theta_h )}  \right)\frac{\left( 3+\cos2 \theta_h  \right) \sin2 \phi_h }{ 8 \sqrt{2}}$
}\\
&\\
 \multirow{2}{*}{$h_+$ (QWP, $f=1/2l$)}   &
    \multirow{2}{*}{
$  \left(1 + \, e^{ i\pi  \cos\theta_h }  \right)\frac{\left( 3+\cos2 \theta_h  \right) \sin2 \phi_h }{ 4\sqrt{2}}$
}\\
&\\
 \multirow{2}{*}{$h_+$ (QWP, $f=1/l$)}   &
    \multirow{2}{*}{
$ \left(1 - \, e^{2i\pi   \cos\theta_h }  \right)\frac{\left( 3+\cos2 \theta_h  \right) \sin2 \phi_h }{ 4 \sqrt{2}}$
}\\
&\\
\hline
    \end{tabular}
    \caption{Response functions for axion DM and GWs when the laser is held on the cavity resonance. }
    \label{table:response_functions}
\end{table}
\begin{figure}[t]
\centering
\includegraphics[width=0.99\textwidth]{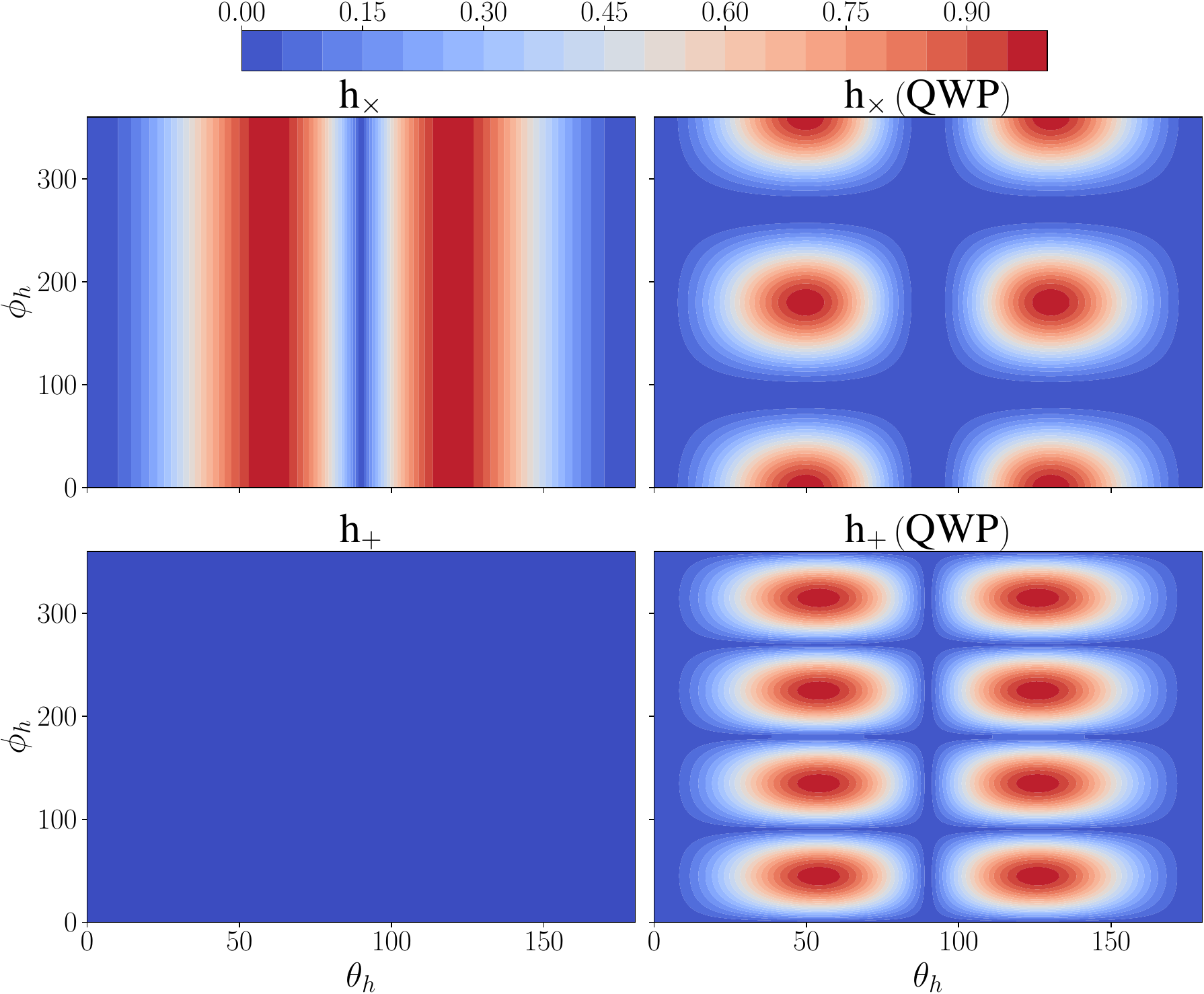} 
\caption{Normalized antenna patterns of the response functions at the second cavity resonance, $f = 1/l$,
when the laser is held on the cavity resonance.
See text and Table~\ref{table:response_functions} for details.
}
\label{fig:gws}
\end{figure}
As ${\cal M}= {\cal O}(s)$, the phase $\phi_n$ in the previous equation can be approximated at zeroth order in $s(f)$. Employing Eq.~\eqref{eq:phaseshift} for this, we find at leading order in $s(f)$ 
\begin{eqnarray}
 \label{eq:cav_resp_matrix0}
 {\bf \hat{s}} ^\dagger \cdot {\bf E}_1 (0)   = &&  t_1|{\bf E}_{\rm in}|\int_{-\infty}^{\infty} df\left(\frac{r_1 r_2 e^{-4 i\pi f_L l  }}{e^{-4 i\pi f_L l  }-r_1 r_2}\right) \left( \frac{1}{e^{-4 i\pi (f_L+ f) l }- r_1 r_2 }  \right)      \\  && \times {\bf \hat{s}} ^\dagger \left( e^{-2i \pi f l  }\int^{l}_{0} \!\! dt' \,   e^{-2 i\pi f t'}\mathcal{P}\tilde{{\cal M}}\mathcal{P}\Bigg|_{{\bf x}(t') = \hat{{\bf z}} \,(l-t') }+    \int^{l}_{0 } \!\! dt' \, e^{-2i \pi f t'}   \tilde{{\cal M}} \Bigg|_{{\bf x}(t') = \hat{{\bf z}} \,t'} \right)   {\bf \hat{p}}. \nonumber
\end{eqnarray}
This motivates to define a \textit{response function due to polarization change} as
\begin{eqnarray}
 \label{eq:cav_resp_matrixdef}
 {\bf \hat{s}} ^\dagger \cdot {\bf E}_{1} (t)  =  t_1|{\bf E}_{\rm in}| \int_{-\infty}^{\infty} df s(f)\,  {\cal H} (f) e^{2 i\pi f t }\,,
\end{eqnarray}
with
\begin{eqnarray}
     \label{eq:cav_resp_matrix}
       \!\!\!\!\!\!\!\!\!\!
s(f) \, {\cal H} (f)   = && {\cal H}_0 (f)  \int^{l}_{0 } \!\! dt' \, e^{-2i\pi f t'}  {\bf \hat{s}} ^\dagger \left( e^{-2i \pi f l  }\mathcal{P} \tilde{{\cal M}}\mathcal{P}\Bigg|_{{\bf x}(t') = \hat{{\bf z}} \,(l-t') }+    \tilde{{\cal M}} \Bigg|_{{\bf x}(t') = \hat{{\bf z}} \,t'} \right)   {\bf \hat{p}} \,,
\end{eqnarray}
and
 \begin{equation}
 {\cal H}_0 (f)=
\left(\frac{r_1 r_2 e^{-4 i\pi f_L l  }}{e^{-4 i\pi f_L l  }-r_1 r_2}\right) \left( \frac{1}{e^{-4 i\pi (f_L+ f) l }- r_1 r_2 }  \right)\,.
 \end{equation}
These response functions are reported in the upper panel of Table~\ref{table:response_functions}.\footnote{For their derivation we use
\begin{align}
{\bf \hat{s}} ^\dagger  \mathcal{P} \tilde{{\cal M}}\mathcal{P}{\bf e}_p\Bigg|_{{\bf x}(t')  = \hat{{\bf z}} \,(l-t') }  \!\!\!\!\! = \epsilon   \times
\left\{
\begin{array}{ll} 
-  (\mathcal{P}{\bf \hat{s}}  \times \mathcal{P}{\bf e}_p)\cdot  
 \mathcal{P}{\bf k}\, &\quad\text{for axions}, \\ 
\frac{i \, \omega\left(-1 + 2 \cos\theta_h\cos^2\phi_h \right) }{\sqrt{2}}
\cos^2\left(\frac{\theta_h}{2}\right) e^{ 2i \pi f{\bf \hat q} \cdot {\bf x}(t')}
& \quad\text{for GWs ($\times$)},
\\ 
 \frac{i \, \omega( 1+\cos^2\theta_h ) \sin2 \phi_h }{2 \sqrt{2}}  \cos^2\left(\frac{\theta_h}{2}\right)  
e^{ 2i \pi f{\bf \hat q} \cdot {\bf x}(t')}
& \quad\text{for GWs ($+$)}. 
\end{array}
\right.
\label{eq:PMPempty}
\end{align}
and
\begin{align}
{\bf \hat{s}} ^\dagger   \tilde{{\cal M}}  {\bf e}_p\Bigg|_{{\bf x}(t') = \hat{{\bf z}} \,t'}   = \epsilon   \times
\left\{
\begin{array}{ll} 
-   ({\bf \hat{s}}  \times {\bf e}_p)\cdot  {\bf k}\,
 &\quad\text{for axions}, \\ 
 \frac{i \, \omega (1 + 2 \cos\theta_h \cos^2\phi_h )}{\sqrt{2}}  \sin^2\left(\frac{\theta_h}{2}\right) e^{ 2i \pi f{\bf \hat q} \cdot {\bf x}(t')}
& \quad\text{for GWs ($\times$)},
\\ 
  \frac{i \, \omega\left(1 + \cos^2\theta_h\right) \sin2\phi_h}{2\sqrt{2}} \sin^2\left(\frac{\theta_h}{2}\right) 
 e^{ 2i \pi f{\bf \hat q} \cdot {\bf x}(t')} \,
& \quad\text{for GWs ($+$)}. 
\end{array}
\right.
\label{eq:Mempty}
\end{align}
} This is a key result of this paper. 
 Before discussing its implications for each specific case, let us point out some general features of ${\cal H}(f)$. 
  The integral in Eq.~\eqref{eq:cav_resp_matrix} gives the relative amplitude of $s$-polarized component after one round trip of the laser through the cavity, whereas the factor ${\cal H}_0 (f)$ accounts for the power build-up of the cavity.

On the one hand, the first factor in  ${\cal H}_0 (f)$ is  independent of the axion or the GW, and is nearly the same enhancement as the one that  $p$ polarization receives,  see Eq.~\eqref{eq:Enobkg}. If the mirrors are tuned so that $2f_Ll  \in \mathbb{Z}$, that is $e^{-4i\pi f_Ll}=1$, 
we find
\begin{align}
 {\cal H}_0 (f)=
\left(\frac{r_1 r_2}{1-r_1 r_2}\right) \left( \frac{1}{e^{-4 i\pi f l }- r_1 r_2 }  \right)\approx  \frac{{\cal F}/\pi}{e^{-4 i\pi f l }- r_1 r_2 } \,, &&\text{with}&&
{\cal F} = \frac{\pi \sqrt{r_1 r_2}}{1-r_1  \, r_2}\,.
 \end{align}
In this expression, we use the fact that the mirrors are designed to be highly reflective such that $1-r_{1,2}\ll1$.  
Hence, the response functions receive a large boost from the finesse  of the cavity, $\mathcal{F}$, which reaches values of the order 100,000 for cavities with lengths of a few hundred meters.  Due to this boost, from now on we will assume that the laser is held on the cavity resonance, that is, we take $e^{-4i\pi f_Ll}=1$.   On the other hand, the second factor in  ${\cal H}_0 (f)$ depends on the sidebands induced by the axion or the GW, namely, $f_L+f$. For instance, for the monochromatic signal in Eq.~\eqref{eq:epsilonmono},  $ {\bf \hat{s}} ^\dagger \cdot {\bf E}_{1}(t)$ will have two terms associated with sidebands at $f_L \pm f_0$. Then, a further resonance  boost is expected if one of these sideband frequencies is an integer multiple of $1/2l$.  

The left panel of Fig.~\ref{fig:gws} illustrates the non-trivial angular dependence of these functions for GWs matching the second resonant frequency. 
 Let us also note that for $\theta_h=0$, the GW response functions do not vanish in general. At first glance, this appears to contradict Ref.~\cite{Pegoraro_1980} (see also Ref.~\cite{Iacopini:1979ww}), which claims that there can be no polarization change if the laser propagates parallel to the GW and no other medium (such as a dielectric) is present. However, upon closer inspection of Eqs.~\eqref{eq:PMPempty} and \eqref{eq:Mempty} indicates that our results actually are in agreement with that: The part of the response function associated with propagation parallel to the GW vanishes, while the term corresponding to propagation antiparallel to the GW does not. Having mirrors reflecting the laser is therefore crucial. 

%

\subsection{Response function with a quarter-wave plate}

To further exploit polarization effects, one might place a  birefringent device, such as a quarter-wave plate (QWP),  between the two mirrors.  For illustration,  we will consider a pair of QWPs each positioned in close proximity to one of the mirrors such that at each of them ${\bf e}\to   {\cal Q}  {\bf e} $, where
\begin{align}
{\cal Q} = \begin{pmatrix}
        1 & 0 & 0 \\
        0 & -i & 0 \\
        0 & 0 & 1
    \end{pmatrix}.
    \label{eq:Q}
\end{align}
Hence, the effect of each QWP is to introduce a relative phase difference of $\pi/2$ between the two polarization components. Note that we assume that both QWPs have the same ${\cal Q}$, i.e., their fast axes are assumed to be aligned to the vertical. Computing the response functions is analogous to the case discussed above. In the absence of an axion or a GW background, the electric field ${\bf E}_1 (0)$ to the right of the QWP near mirror 1 is also given by Eq.~\eqref{eq:eansatz}. However, instead of Eq.~\eqref{eq:eansat_vacuum}, here we have 
\begin{align}
    L_j =   r_1 
    \mathcal{P}' \times r_2 \mathcal{P'}  \,, && \text{with} &&\cal {P}' = {\cal Q} \, { \cal P }\, {\cal Q}\,.  
\end{align}
Clearly, the matrix ${\cal P}'$ satisfies Eq.~\eqref{eq:Pproperty}, as it describes the reflection off a QWP, i.e. the effect of light passing through a QWP, being reflected, and then passing through the QWP again. Applying this process twice simply yields the identity. Consequently,  we have $L_j = r_1 r_2 \mathbb{1}$, implying that the presence of two QWPs does not affect the result of Eq.~\eqref{eq:Enobkg}. In particular, the laser remains  $p-$polarized.  
Similarly, in presence of an axion or a GW background, instead of Eq.~\eqref{eq:master_form_L}, we now have 
\begin{eqnarray}
    L_j &=& \underbrace{\cal Q}_{\text{QWP}}\, \,\underbrace{r_1  
    \mathcal{P} }_{\text{Reflection off mirror 1}} \, \underbrace{\cal Q}_{\text{QWP}} \times  \underbrace{\left( \left(\cdots\right)\mathbb{I} + \int^{l}_{0}  dt' \,  {\cal M} (t'-(2 j\!+\!1)l)\Bigg|_{{\bf x}(t') = \hat{{\bf z}} \,(l-t') } \right)}_{\text{ Path from mirror 2 to mirror 1
} }  \nonumber \\ &\times& \! \!  \underbrace{\cal Q}_{\text{QWP}} \, \underbrace{r_2 \mathcal{P} \, 
}_{\text{Reflection off mirror 2}} \, \underbrace{\cal Q}_{\text{QWP}} \,  \times \underbrace{\left(\left(\cdots\right)\mathbb{I} + \int^{l}_{0 }  dt'  {\cal M} (t' -  (2j\!+\!2) l)\Bigg|_{{\bf x}(t') = \hat{{\bf z}} \,t' } \right)}_{\text{ Path from mirror 1 to mirror 2}}  .
\label{eq:master_form_L_QWP}
\end{eqnarray}
The response functions with QWP can be obtained from those without the QWP by replacing ${\cal P} \to {\cal P}'$. Plugging\footnote{In more detail, instead of Eq.~\eqref{eq:PMPempty}, we now have
\begin{align}
{\bf \hat{s}} ^\dagger  \mathcal{P'} \tilde{{\cal M}}\mathcal{P'}{\bf e}_p\Bigg|_{{\bf x}(t')  = \hat{{\bf z}} \,(l-t') }  \!\!\!\!\! = \epsilon   \times
\left\{
\begin{array}{ll} 
-  (\mathcal{P'}{\bf \hat{s}}  \times \mathcal{P'}{\bf e}_p)\cdot  
 \mathcal{P'}{\bf k} = -   ({\bf \hat{s}}  \times {\bf e}_p)\cdot  {\bf k} \, &\quad\text{for axions}, \\ 
-\frac{i \, \omega\left(-1 + 2 \cos\theta_h\cos^2\phi_h \right) }{\sqrt{2}}
\cos^2\left(\frac{\theta_h}{2}\right)  e^{ 2i \pi f{\bf \hat q} \cdot {\bf x}(t')}
& \quad\text{for GWs ($\times$)},
\\ 
 -\frac{i \, \omega( 1+\cos^2\theta_h ) \sin2 \phi_h }{2 \sqrt{2}}  \cos^2\left(\frac{\theta_h}{2}\right)  
 e^{ 2i \pi f{\bf \hat q} \cdot {\bf x}(t')}
& \quad\text{for GWs ($+$)}. 
\end{array}
\right.
\label{eq:PpMPpempty}
\end{align}} this expression in Eq.~\eqref{eq:cav_resp_matrix}, we obtain the results in the lower panel of  Table~\ref{table:response_functions}. Several comments are in order. The axion response function without QWP exactly matches the one reported in Ref.~\cite{Nagano:2019rbw}, calculated by a slightly different method.
Table \ref{table:response_functions} shows that the suppression in this response function arises when $f\ll 1/l$  can be avoided by placing the QWPs. This is particularly useful for axion DM because, as we will see, this improves the sensitivity in the low mass range at the cost of suppressing the resonant peaks at higher frequencies. 

Likewise, the QWPs are also instrumental for GWs with $+$ polarization. Without them, due to the multiplicative factor,  $1-e^{-4i \pi f l}$, the response function is suppressed with respect to that of $\times$ polarization when the sidebands are in resonance.\footnote{Here we should emphasize that the definitions of $+$ and $\times$ polarizations given in Eq.~\eqref{eq:planewave} are coordinate dependent. The coordinate independent statement is that only a single combination of the 
GW polarizations will be unsuppressed for a given polarization state of the laser. See Ref.~\cite{Domcke:2023bat} for a related discussion in the context of axion haloscopes.} This is also evident in Fig.~\ref{fig:gws}.
This is reminiscent of the coupling of GWs to toroidal axion haloscopes, where, in a similar fashion, the signal associated with $+$ mode is suppressed with respect to  the $\times$ polarization when the readout device is circular. As discussed in Refs.~\cite{Domcke:2022rgu,Domcke:2023bat}, this stems from selection rules arising from symmetry considerations.  Although the suppression occurs only when the sidebands are in resonance, the method proposed here is most effective in that regime; otherwise, the polarization rotation is of the order of the GW amplitude\footnote{This statement remains valid even beyond the geometric limit (see, e.g., \cite{Domcke:2024abs}).}. This motivates the use of a resonant cavity to harness resonant boosts and the QWPs to eliminate suppressions.

\subsection{Realistic setups}
\begin{figure}[t]
    \centering
    \includegraphics[width=0.99\textwidth]{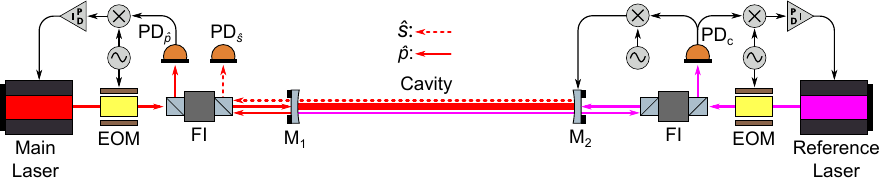}
    \caption{Schematic of the experimental setup. Here the main laser is injected to the cavity through mirror $\rm M_1$. The $s$ and $p$ polarization states, represented by solid and dashed lines, are split in reflection of the cavity by a FI. The $p$-polarized light is used to stabilize the frequency of the main laser to the cavity resonance using PDH laser frequency stabilization. This also necessitates the EOM  directly after the laser. The axion or GW signal is measured at $\rm PD_s$. On the right side of the figure a reference laser is injected to the cavity. This laser is also frequency stabilized to one of the resonances using the PDH technique. In addition, the interference beatnote between the main laser and reference laser measured at $\rm PD_c$ is used as the error signal in a feedback control loop to stabilize the absolute length of the cavity.} 
    \label{fig:detectorpic}
\end{figure}

A few modifications to the arrangement of Fig.~\ref{fig:ang_plot}  will enable us to apply our formalism to a realistic experimental setup. For concreteness, 
we propose the setup depicted in Fig.~\ref{fig:detectorpic}, which consists of a cavity equipped with a detection system capable of measuring the polarization change induced by the background of axions or GWs. On the left side of the diagram an electro-optic modulator (EOM) is used to induce phase modulation sidebands on the main laser, so that the Pound-Drever-Hall (PDH) frequency stabilization technique can be used to hold its frequency on resonance with the cavity \cite{drever1983laser,black2001introduction}. The laser then passes through a Faraday Isolator (FI) before being incident on the cavity. In reflection of the cavity, the $p$-polarized light is reflected by the port on the left side of the FI and is then incident on the photodetector (PD) $\rm PD_p$, which is used as the sensor for the PDH frequency laser stabilization system. The $s$-polarized light in reflection of the cavity is reflected at the port on the right side of the FI and is incident on $\rm PD_s$. This is the PD used to sense the signal associated with GWs or axion DM. On the right side of the diagram is the length control system for the cavity. Here a reference laser is also frequency stabilized to the cavity via the PDH technique, but at a different FSR from the main laser.\footnote{We should acknowledge that at this specific resonance the reference laser will create a background causing a loss in sensitivity at that point, however the frequency of the reference laser can be tuned to be outside the range we are exploring without inhibiting the length stabilization system.} The interference beatnote between the two lasers is then incident on $\rm PD_c$.
The phase of the interference beatnote can be sensed by mixing the electronic signal from $\rm PD_c$ with a stable oscillator thus revealing the length changes in the cavity. This signal can also be used to generate an error signal for a feedback control loop that can maintain the absolute length of the cavity at a fixed value. This is a critical system in the experiment as if the length of the cavity is left free-running, the frequency associated with the resonances in ${\cal H}(f)$ will constantly be changing with respect to $f$.

Under these conditions, the electric field at $\rm PD_s$ is given by \begin{equation}
{\bf E}_{\rm PD}(t) =  \, \left[   t_1^2 |{\bf E}_{\rm in}|  \left( \alpha +  \int_{-\infty}^{\infty} df s(f) \,  {\cal H} (f) e^{2i\pi f  t} \right) + { \rm E}_{\rm noise}(t) \right]  e^{-2i\pi f_L t} {\bf \hat{s}} .
\label{eq:EDP0}
\end{equation}
    The term proportional to $\alpha$ accounts for the intrinsic  birefringence of the cavity as the cavity mirrors themselves will have a slight intrinsic birefringence leading to small additional complex terms in both the diagonal and non-diagonal elements of $\cal P$. This will lead to the accumulation of a minor static $s$-polarized component of the light circulating in the cavity such that $\frac{{\bf \hat{s}} ^\dagger \cdot {\bf E}_1}{|{\bf E}_1|} =\alpha$ with $\alpha\ll1$. 
Ref.~\cite{Nagano:2021kwx} considered a similar setup for axions: they use its birefringence, $\alpha$, in reflection of the cavity, and a half-wave plate  in transmission of the cavity, that is, an additional birefringence on top of the one associated with axions. The second term in Eq.~\eqref{eq:EDP0} is the polarization change induced by the background of axions or GWs as encoded in the response function of Eq.~\eqref{eq:cav_resp_matrix}. Finally, ${\rm E}_{\rm noise}(t)$ is a generic term for the noise, which can originate from various sources and will be quantified  below. We assume that the  term in Eq.~\eqref{eq:EDP0} proportional to $\alpha$ dominates over the rest. 
The virtue of this is that the signal of the background of axions or GWs appears as an amplitude modulation of the one induced intrinsically by the cavity. Concretely, at the detection port  the  power is determined by\footnote{Strictly speaking, in the case of a GW background, to calculate $|{\bf E}_{\rm PD}|^2$ we should use $(\eta_{ij}+h_{ij})E^i_{\rm PD}E^{j*}_{\rm PD}$. Nevertheless this extra contribution is negligible with respect to the one we account for, which is multiplied by the resonant enhancement factors in ${\cal H}(f)$. }
\begin{eqnarray}  
    \label{eq:power}
    \!\!\! \!\!\! \!\!\!
 \left| {\bf E}_{\mathrm{PD}}\right|^2 
 &\simeq& \, \underbrace{\alpha^2 t_1^4 |{\bf E}_{\rm in}|^2  }_{\sim P_{\rm out}}  + \underbrace{2   |{\bf E}_{\rm in}|^2  \alpha\, t_1^4 \int_{-\infty}^{\infty} df s(f) \,  {\cal H} (f) e^{2 i \pi f t}}_{\sim P_{\rm signal}}  + \underbrace{2 \alpha \, t_1^2  \, |{\bf E}_{\rm in}|{\rm E}_{\rm noise}(t)}_{\sim P_{\rm noise}},
    \end{eqnarray}
where we  expanded at linear order in $s(f)$ and $E_{\rm noise}$.
Therefore, the power arriving at the detector consists of three contributions: $P_{\rm out}$, representing the ordinary power of the laser beam; $P_{\rm signal}$, corresponding to the contribution from the background of axions or GWs; and $P_{\rm noise}$, representing the noise contribution that limits the sensitivity.

It is important to notice that the term proportional to $\alpha$  allows the signal associated with axions or GWs to be linear in $s(f)$, rather than quadratic. This is because the induced $s$-polarized mode modulates the already existing $s$-polarized mode induced by the cavity alone. 
According to Eq.~\eqref{eq:power}, this linear relation is given by the following transfer function
\begin{align}
P_{\rm signal}(f) ={\cal T}(f) s(f)\,, && \text{where} && 
{\cal T}(f)=  \frac{2P_{\rm out}}{\alpha}   {\cal H} (f) \,.
\end{align}
As is commonly done, this enables us to define the noise spectral density as
\begin{equation}
\langle P_{\rm noise} (t)P_{\rm noise} (t')\rangle =\frac{1}{2}\int^\infty_{-\infty} df S^{\rm noise }(f) | {\cal T}(f)|^2 e^{-2i\pi f (t-t')}\,. 
\label{eq:noise_spectral_density_def}
\end{equation}
Having introduced $ S^{\rm noise }(f)$, we can now discuss prospects at ALPS II. 
\section{Sensitivity prospects using ALPS\,II cavities} 
\label{sec:ALPSII}

With ALPS\,II scheduled to complete data taking in the next few years, it may be soon be possible to use the optical system for polarimetric searches for axions and GWs. For this reason we consider cavities designs, reported in Table~\ref{tab:values}, that would already work with the existing ALPS\,II infrastructure and could be adapted to measure these polarization effects, following the setup proposed above.  Specifically, we investigate cavity lengths of $l=245$\,m and $l=20$\,m.
The corresponding first resonant peak is at $f = 1/2l \simeq 612 \, {\rm kHz}  \simeq 2.5\, {\rm neV}$.  
In this frequency range, the primary source of noise is  the quantum fluctuations of the laser, i.e., shot noise. It is possible that the dynamic birefringence noise intrinsic to the cavity also plays a role in the sensitivity, however this is not expected to be the case for frequencies above 1\,kHz \cite{hartman2019characterization}.
Other sources of noise traditionally associated with interferometric GW observatories such as radiation pressure and seismic noise, are not relevant over the frequency range considered here\footnote{It should be noted here that the absolute length of the cavity must be controlled such that ${\cal H}(f)$ is stable over the duration of the measurement to ensure optimal sensitivity at the resonances.}.

Quantum fluctuations associated with shot noise are determined by a Poisson distribution, $\sqrt{\langle P_{\rm noise} (t)^2 \rangle}/P_{\rm out} = 1/\sqrt{N_\gamma}$,  and are not correlated at different times. Hence
\begin{equation}
\frac{1}{P_{\rm out}^2}\langle P_{\rm noise} (t)P_{\rm noise} (t')\rangle \bigg|_{\rm shot 
\,noise} =\frac{T_{\rm obs}}{N_\gamma} \delta(t-t')\,.
\end{equation}

In the absence of a signal (i.e axion DM or GWs), the number of photons arriving at  $\rm PD_s$  during a time $T_{\rm obs}$ is determined by the power as  
$N_{\gamma} = P_{\rm out} \, T_{\rm obs}/ \omega_L  $, or according to Eq.~\eqref{eq:power}, $N_{\gamma}  = \alpha^2 t_1^4 P_{0} \, T_{\rm obs}/ \omega_L$, where  $P_0$ is the laser power injected in the cavity. Then, Eq.~\eqref{eq:noise_spectral_density_def} gives the noise spectral density as 
\begin{equation}
\label{eq:Snoise}
\left(S_{\mathrm{noise}} (f)\right)^{1/2} = \frac{1}{t_1^2|{\cal H}(f)|} \sqrt{\frac{\omega_L}{2P_0}}, 
\end{equation}
where $t_1^2 |{\cal H}(f)| \sim {\cal F}/\pi$ at the resonance peaks.
The quantity $P_0$   can be computed from the maximum possible circulating power inside the cavity, $P_{\rm max}$, reported in Table~\ref{tab:values}. From Eq.~\eqref{eq:Enobkg} and considering the  field, we have
\begin{equation}
P_0 = P_{\rm max} \left(\frac{1-r_1 r_2}{t_1}\right)^2. 
\end{equation}
The chosen values for $P_{\rm max}$ are consistent with what has been achieved in ALPS\,II \cite{Spector:2019ooq}. Here the primary limitation on the power is believed to be related to absorption effects in the cavity mirror coatings, which impose a maximum intensity that the cavity can tolerate. For this reason, the longer cavity with larger beam spots sizes on the mirrors should be capable of operating with a higher circulating power.\footnote{Here we assume a near confocal geometry for the cavities to minimize the beam spot sizes on the mirrors, thus reducing clipping losses in the beam tube and scattering losses on the surface of the mirrors. This does have the effect of potential limiting the maximum circulating power though, by increasing the relative peak intensity of the field at the mirrors.} We should note that while the optimal sensitivity at the resonances is still achieved when the cavity is impedance matched \cite{bond2016interferometer}, i.e. $t_1^2=t_2^2+{\ell}^2$, once the cavity has a sufficiently high finesse to achieve $P_{\rm max}$ for a given maximum input laser power, the gain in sensitivity from increasing the cavity finesse will only be proportional to the square root of the increase in finesse.

\begin{table}[t]
    \centering 
    \begin{tabular}{|llcl|c|c|c|}
\hline
\hline
QWP   & $l \, [\mathrm{m}]$  & $P_{\mathrm{max}}  \, [\mathrm{kW}]$  &$(t^2_1, t^2_2, {\ell}^2)$ (ppm) & ${\cal F}$ & $m_{\rm max}$ (eV) & $\tau_{\rm storage}$ (ms)
   \\
\hline
No & $245$  & $150 $ &   $(22,2,20) $  & $1.4 \times 10^5$ & $4.2\times10^{-8}$ &
74.2\\
Yes & $245$  & $10$ & $(1100,100, 1000) $ &  $2.9\times 10^3$ & $2.1\times10^{-6}$ 
&
1.48\\
No & $20$  & $50 $ & $(11, 1, 10)$ & $2.9 \times10^5$ & $2.6\times10^{-7}$
&
12.1\\
Yes & $20$  &$10$  & $(1100,100, 1000)$ & $2.9\times 10^3$ & $2.6\times10^{-5}$
&  0.121\\
\hline
\hline
\end{tabular}
\caption{
Properties of the cavities studied in this work, all utilizing a laser with a wavelength $\lambda = 1064$ nm. See text for details. 
}
\label{tab:values}
\end{table}

Lastly, an important assumption we need to make is that the signal, whether from axion DM or GWs, must have a coherence time, $\tau$, longer than the storage time of the cavity, $\tau_{\rm storage}$, defined as the time in which the cavity field decays by $1/e$ \cite{isogai2013loss}. Therefore, we must impose the following condition
\begin{equation}
\tau > \tau_{\mathrm{storage}} \simeq \frac{2\mathcal{F}}{\pi} \frac{l}{c}\,. 
\label{eq:storage_time}
\end{equation}
Here we define the cavity storage time as the time taken for the cavity field to decay by $1/e$ when no injected field is present.

\begin{figure}[t]
\centering
\includegraphics[width=0.78\textwidth] {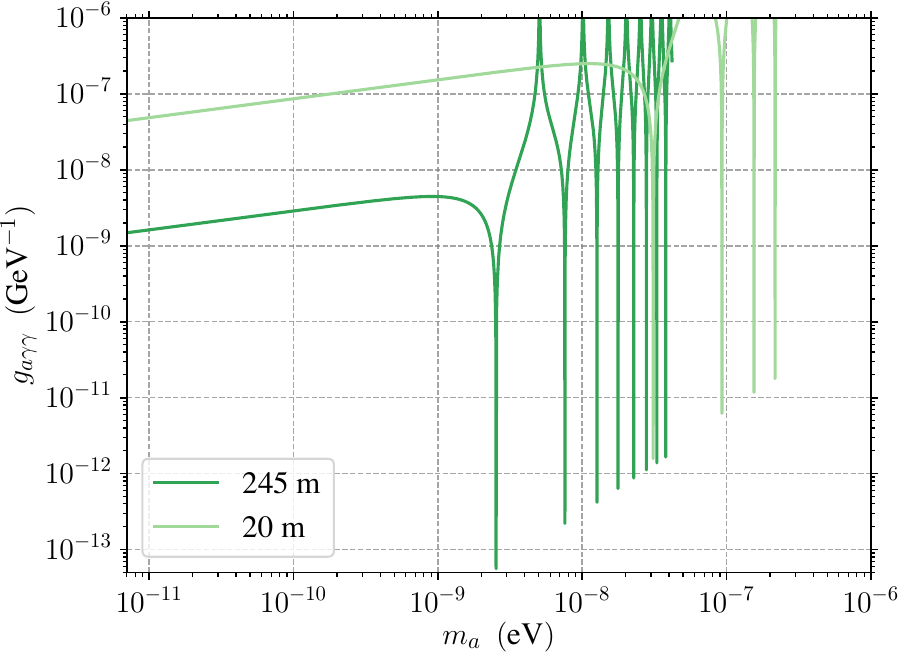} 
\includegraphics[width=0.78\textwidth]{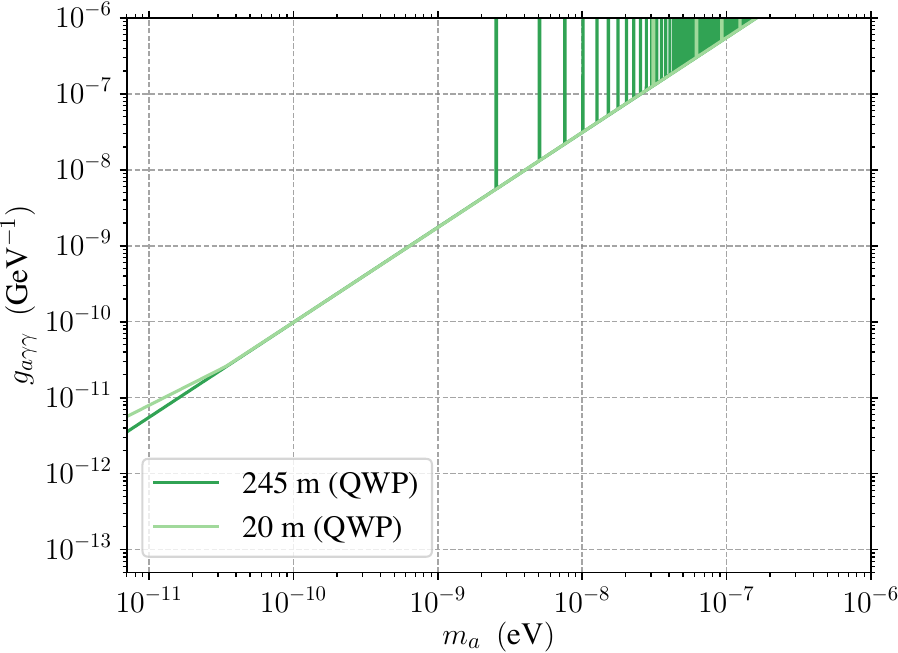} 
\caption{Projected sensitivities to the axion-photon coupling for the different cavity configurations of Table~\ref{tab:values}, assuming an observation time of 30 days. The lines are cut off at  $m_{\rm max}$.}
\label{fig:ax_sens1}
\end{figure}

%
\paragraph{ \bf Sensitivity to the axion-photon coupling. }
In the frequency range relevant for the cavities considered here, the axion coherence time of Eq.~\eqref{eq:axiontau} is always less than the observation time, which we take as $T_{\rm obs}=30$ days. For such a monochromatic signal and coherence time, the signal-to-noise ratio is given by~\cite{Maggiore:2007ulw}
 \begin{align} 
\frac{S}{N} =\frac{\left(T_{\mathrm{obs}} \tau\right)^{1/4}}{\left(S^{\mathrm{noise}}\right)^{1/2}} s_0 
\quad \mathrm{with} \quad  T_{\mathrm{obs}} > \tau. 
 \label{Eq:SNR}
 \end{align}
\begin{figure}[t]
\centering
\includegraphics[width=0.78\textwidth]{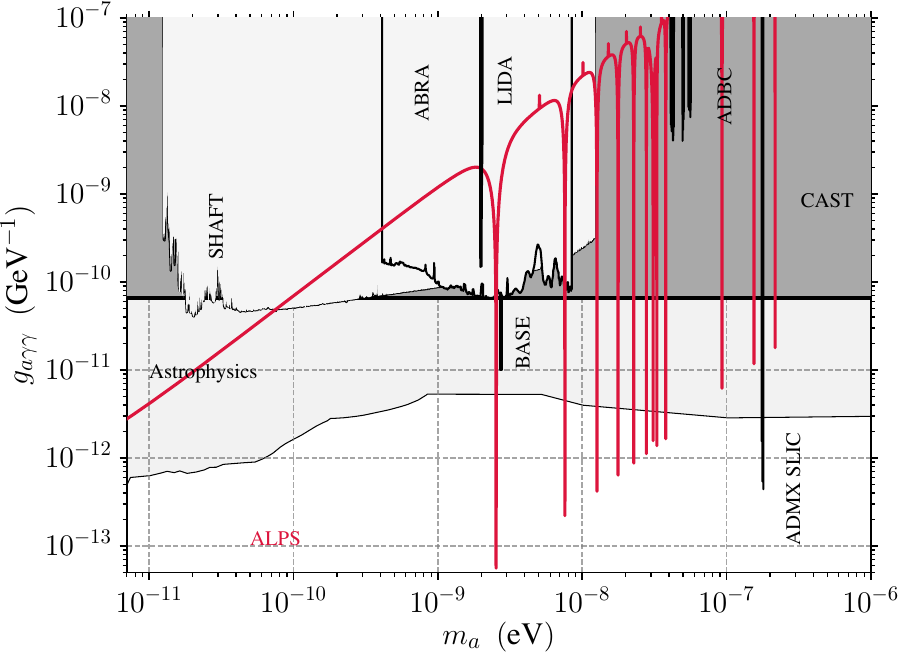}
\caption{%
Projected sensitivity to the axion-photon coupling using the combined projections of the 20\,m and 245\,m cavities with and without QWP, for an observation time of 30 days. This is compared against existing experimental bounds. See text for details.} \label{fig:ax_sens2}
\end{figure}
Here $s_0=\delta c_0$ and, as Eq.~\eqref{eq:defc} shows, this amplitude is proportional to the axion-photon coupling. The value leading to $S/N=1$ is
\begin{equation}
g_{a \gamma \gamma} = \frac{\sqrt2 \omega_L}{\sqrt{ \rho_a}} \frac{(S^{\rm noise}(\frac{m_a}{2\pi}))^{1/2}}{(T_{\rm obs} \, \tau)^{1/4}} \,  \,.
\end{equation}
In Fig.~\ref{fig:ax_sens1}  we show the corresponding projected sensitivity for the different cavities of Table~\ref{tab:values}. Here, the storage time of the cavity, given in Eq.~\eqref{eq:master_form_L} and reported in Table~\ref{tab:values}, sets an upper limit on the mass range, as follows from Eq.~\eqref{eq:storage_time}. For the 245\,m  (20\,m) empty cavity, this corresponds to a mass of 42\,neV (260\,neV) which is why the sensitivity plots have been cut off at those points. The cavities using QWPs have a much lower finesse and are therefore able to probe higher masses, albeit at a lower sensitivity. Concretely, the 245\,m and 20\,m cavities using the QWPs have a maximum mass reach of 2.1\,{\textmu}eV and 26\,{\textmu}eV respectively,  but these masses are not shown on the plots as the corresponding sensitivity is well above existing exclusion limits. We  do not consider masses lower than $10^{-11} \, \mathrm{eV}$ either because, in that frequency range, the dynamic birefringence noise intrinsic to the cavity mirror coatings is expected to limit the sensitivity of the experiment~\cite{hartman2019characterization}.
 
As is clear from these plots, the 245-meter cavity filled with QWPs, along with the empty 245-meter and 20-meter cavities, are complementary and can achieve optimal sensitivities across a wide range of masses.
In Fig.~\ref{fig:ax_sens2}, we present the projected sensitivity obtained by combining the observations from the various cavities in quadrature, assuming 30 days of observation for each one. 
Moreover, we compare the sensitivity of these cavities with existing bounds from other experiments that utilize similar birefringent effects, such as ADBC, BASE, DANCE, and LIDA \cite{Heinze:2023nfb, Pandey:2024dcd, Heinze:2024bdc, Gottel:2024cfj, Oshima:2023csb}, as well as other laboratory experiments like CAST \cite{CAST:2017uph}, ABRACADABRA \cite{Ouellet:2018beu}, SHAFT \cite{Gramolin:2020ict}, and ADMX-SLIC \cite{Crisosto:2019fcj}, and compare these bounds with those from astrophysical observations. Thus, our approach offers competitive limits, particularly for masses matching the resonance frequencies. Furthermore, while we only discuss the sensitivity to axions here, this analysis can be generally applied to ultra-light bosonic dark matter candidates, see e.g. \cite{PhysRevD.107.083035}.

\paragraph{Strain sensitivity for a coherent GW signal.}

\begin{figure}[t]
\centering
\includegraphics[width=0.78\textwidth]{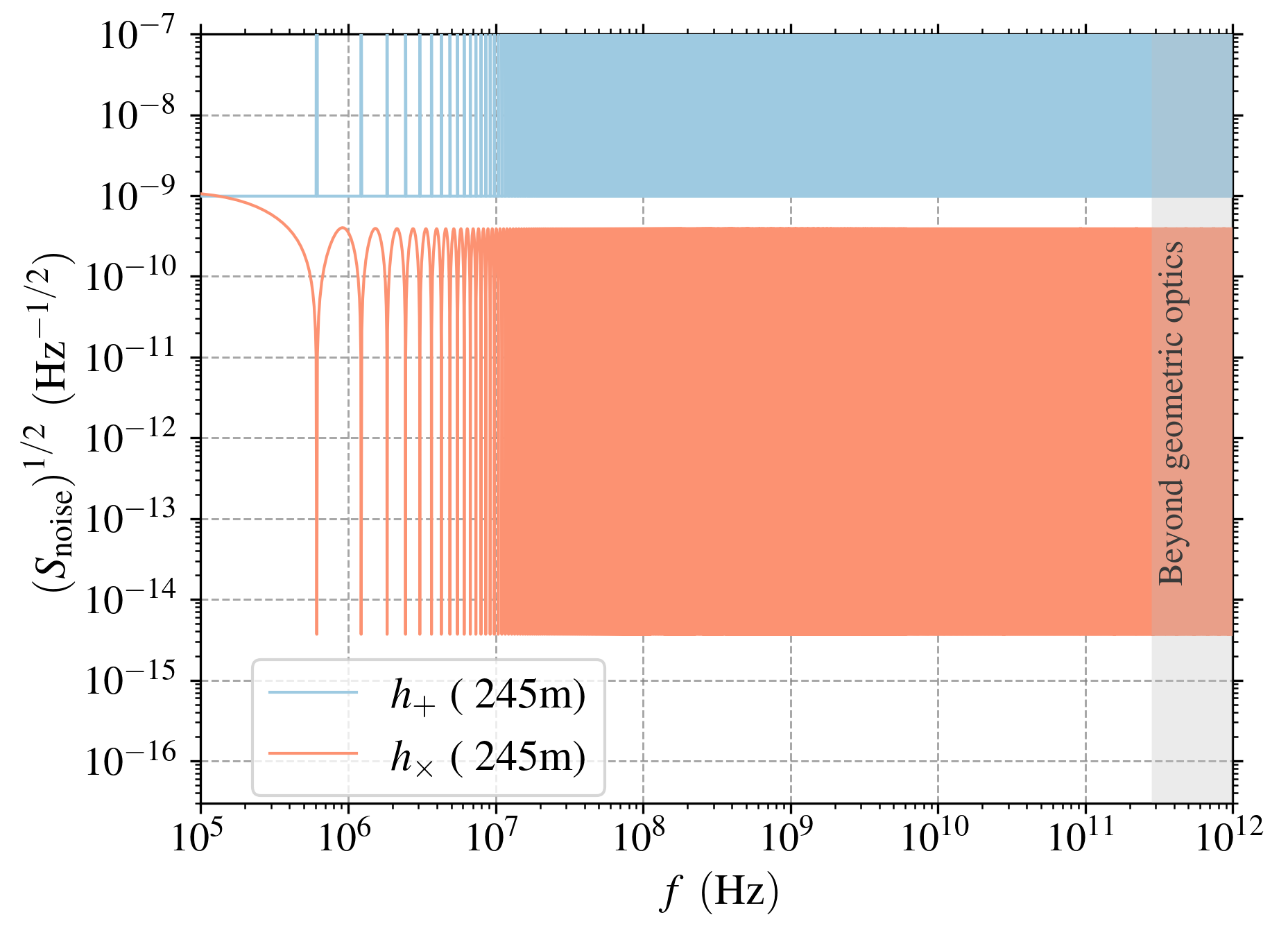} 
\includegraphics[width=0.78\textwidth]{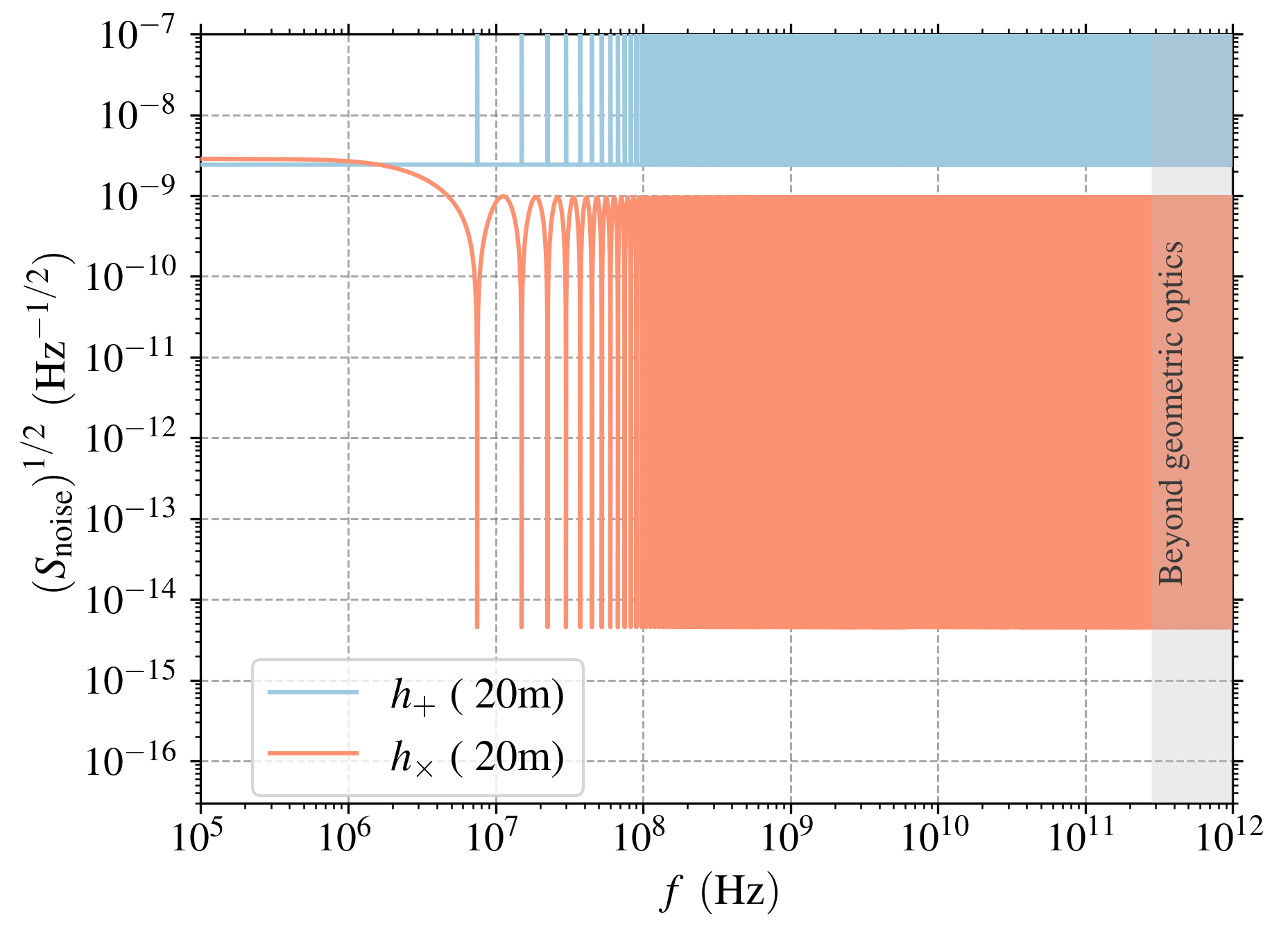} 
\caption{Projected strain sensitivity for the empty cavity configurations of Table~\ref{tab:values}: 245m cavity (\textit{top}) and 20m cavity (\textit{bottom}).
We show the projected sensitivities for the $+$ polarization in blue and for the $\times$ polarization in red.
}  
\label{fig:gw_sens}
\end{figure}

\begin{figure}[t]
\centering
\includegraphics[width=0.78\textwidth]{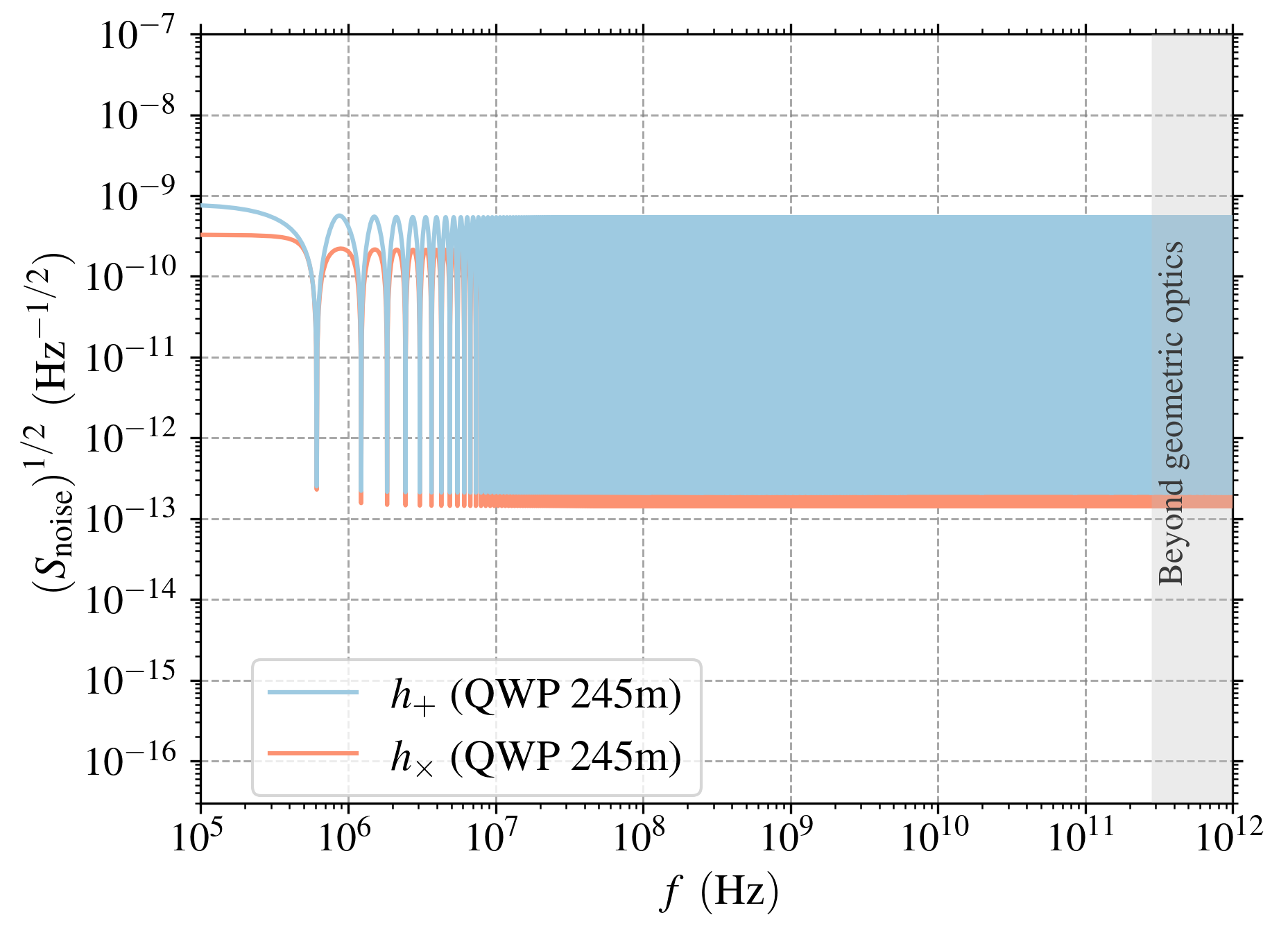} 
\includegraphics[width=0.78\textwidth]{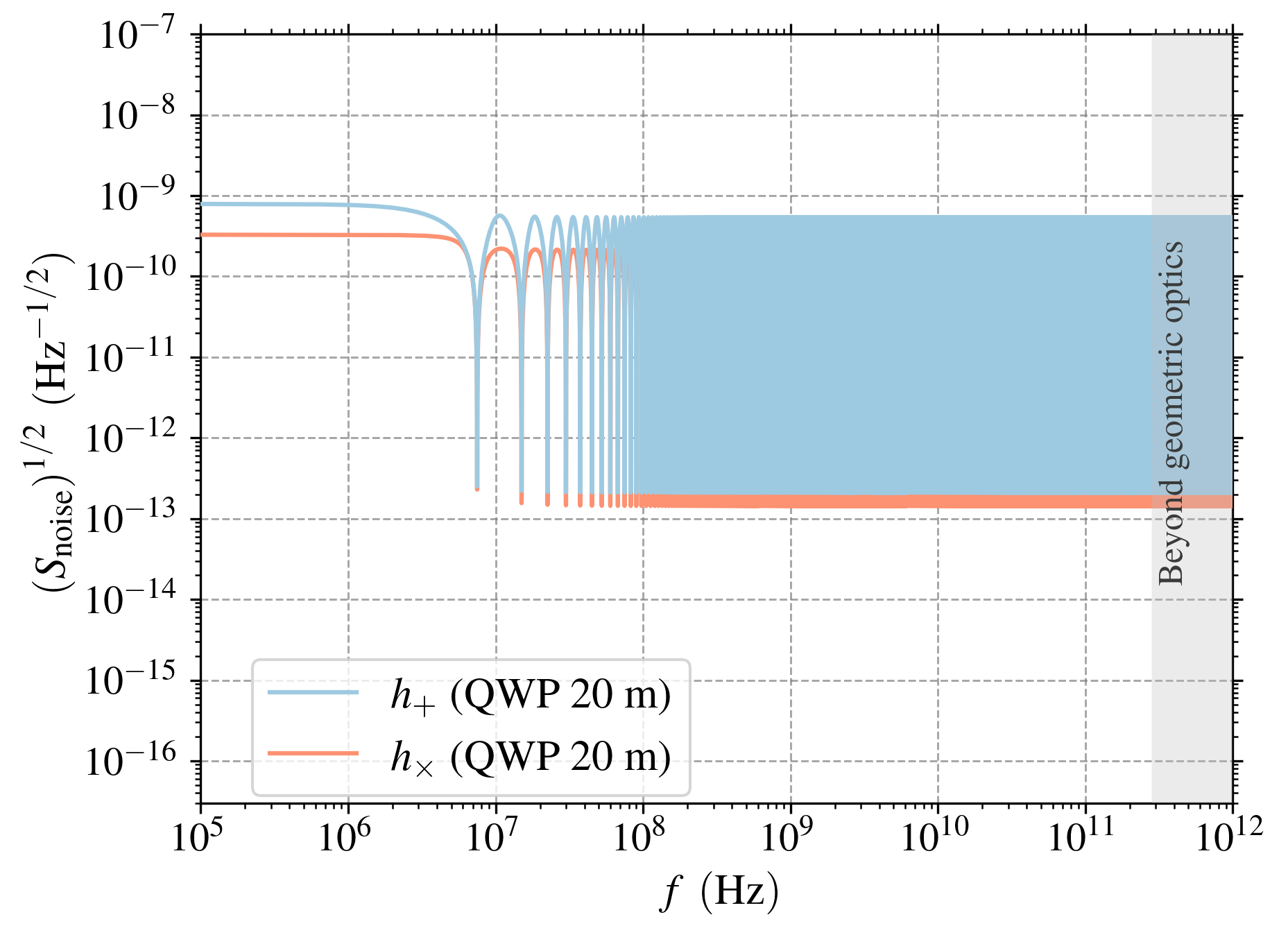}
\caption{Projected strain sensitivity for the cavity configurations with QWPs of Table~\ref{tab:values}: 245m cavity (\textit{top}) and 20m cavity (\textit{bottom}).
We show the projected sensitivities for the $+$ polarization in blue and for the $\times$ polarization in red.
}  
\label{fig:gw_sens0}
\end{figure}

For GWs, we assume a deterministic signal with a coherence time lasting longer than the storage time of the cavity. In Figs.~\ref{fig:gw_sens} and \ref{fig:gw_sens0}, we show the strain sensitivity equivalent to shot noise, obtained in Eq.~\eqref{eq:Snoise}. 
Since the sensitivity depends on the angle of incidence, rather than fixing a particular direction, in Eq.~\eqref{eq:Snoise} we consider the expectation value of the sensitivity over the full sky by taking the following average  %
\begin{equation}
  { \bar {\cal H}}(f) = \left( \frac{1}{4 \pi} \int d\cos{\theta_h} d\phi_h \, |{\cal H}(f; \theta_h, \phi_h)|^2\right)^{1/2}.
\end{equation}
The maximum projected sensitivity in the plots remains constant at high frequencies\footnote{The experiment loses sensitivity to the axion-photon coupling at high frequencies because the amplitude of the axion field is constrained by the DM relic density as  $a_0 \propto 1/m_a$, see Eq.~\eqref{Eq:Ax_A}. No such constraint is assumed for the GW amplitude, which explains the different behavior at high frequencies observed for axions and GWs in Figs.~\ref{fig:ax_sens1}  and \ref{fig:gw_sens}.}
until it reaches the limit of applicability of geometric optics. We conservatively define this limit as GW wavelengths shorter than $10^3$ times the laser wavelength, as shown in the figures. We should also note that measurements at these frequencies, or for that matter over the frequency band we discuss in this paper, will require a sophisticated system to perform the data acquisition, the processing, and analysis, due to the extremely high data rates.

In contrast to the case of axions,  odd and even resonances  give a large sensitivity after averaging over all the possible GW directions. This comes about due to the dependence of ${\cal H}(f)$ on the sky position, as there are some angles in which resonant effects still occur for the even resonances, see Table~\ref{table:response_functions}. Another notable feature of Figs.~\ref{fig:gw_sens} and \ref{fig:gw_sens0} is the lack of dependence of the peak sensitivity on the length of the cavity or the frequency of the resonances.  This is particularly interesting because it contrasts with GW observatories sensing the length changes in the arms of an interferometer. For these interferometric detectors, their peak sensitivity is of course related to the arm length, but they also show a reduced all-sky response at the higher frequency resonances of the cavity. Analyses of the response show this reduction is inversely proportional to the frequency of the resonance being considered \cite{Schnabel:2024hem}. 

\begin{figure}[t!]
\centering
\includegraphics[width=0.82\textwidth]{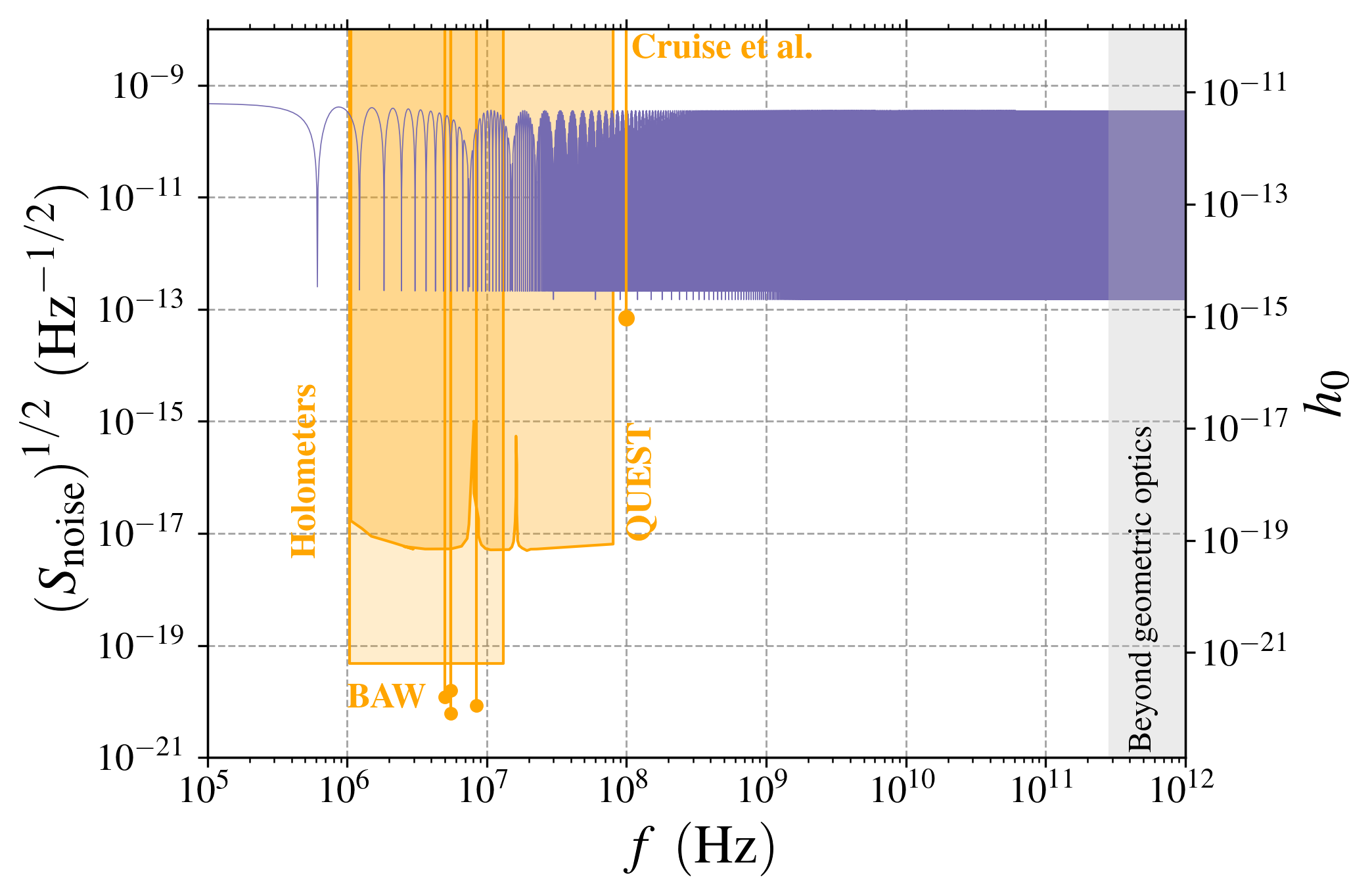} 
\includegraphics[width=0.82\textwidth]{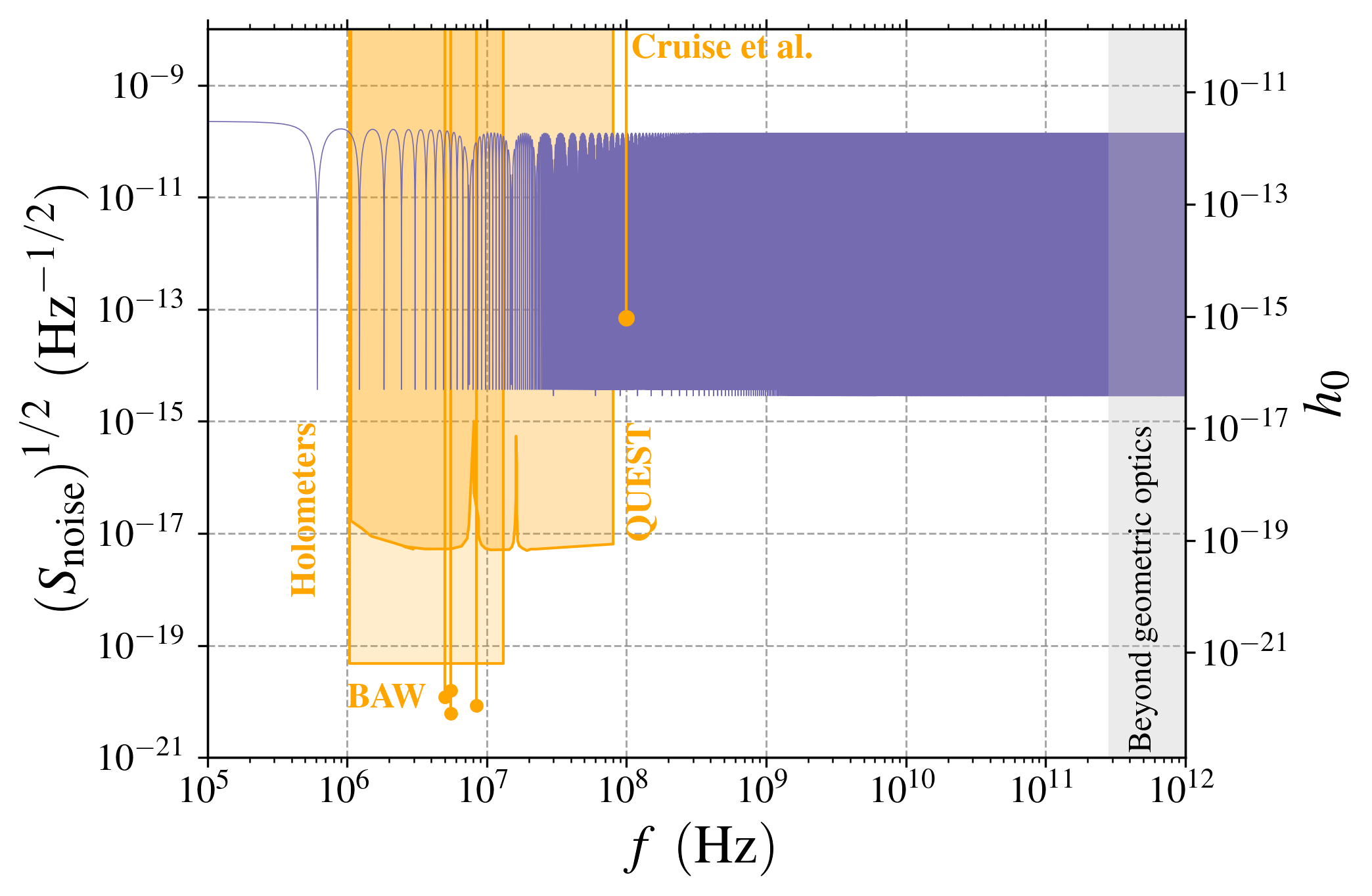} 
\caption{Overall projected strain sensitivities for  the $+$ polarization 
 (\textit{top}) and for the $\times$ polarization (\textit{bottom}) combining in quadrature the the different cavities. The left axis shows the shot noise equivalent strain sensitivity while the right axis shows the sensitivity to $h_0$ assuming a signal with $T_{\rm obs}={1\rm\,year}$ and  $\tau={1\rm\,s}$ as given by Eq.~\eqref{Eq:SNR}. We compare with bounds from bulk acoustic wave oscillations (BAW)~\cite{Goryachev:2014nna, Campbell:2023qbf},  the Fermilab Holometer~\cite{Holometer:2016qoh}, QUEST~\cite{Patra:2024thm}, as well as precision polarimetry measurements performed on a microwave cavity by {\it Cruise, et al}~\cite{Cruise:2006zt}. }
\label{fig:h0_gw}
\end{figure}

Finally, in the left axis of the plots in Fig.~\ref{fig:h0_gw}, we show the projected sensitivity on the strain, considering  observations from the different configurations and combining them in quadrature, as in the case of axion DM. Simultaneously, we show on the right axis the sensitivity on $h_0$ assuming a signal with $T_{\rm obs}={1\rm\,year}$ and  $\tau={1\rm\,s}$ as given by 
Eq.~\eqref{Eq:SNR}, with $s_0=h_0$ and taking $S/N=1$. 
These projections are compared with existing experimental results in this frequency regime, including those utilizing bulk acoustic wave oscillations (BAW) \cite{Goryachev:2014nna, Campbell:2023qbf}, the Fermilab Holometer \cite{Holometer:2016qoh}, QUEST \cite{Patra:2024thm}, as well as precision polarimetry measurements performed on a microwave cavity by {\it Cruise, et al} \cite{Cruise:2006zt}. For the Holometer and QUEST we plot the single interferometer linear spectral density of the strain sensitivity rather than the strain sensitivity in terms of the cross spectral density. The reason for this is that here we only consider deterministic signals instead of stochastic signals.

We note that the methods described in this paper could be compatible with sensing techniques that use cross spectral densities to help identify signals and reduce the systematic uncertainty of the experiment. One of these techniques, discussed in Ref.~\cite{Venneberg:22}, makes use of quantum correlation measurements (QCM) to measure the power noise spectrum of a laser a factor of ten below the shot noise limit in the frequency region we are considering. It is important to point out that using QCM or other techniques that make use of cross spectral densities will not improve the sensitivity of the experiment to  deterministic signals relative to the theoretical shot noise limit (as opposed to that of stochastic signals). In practice though, these techniques may be important to help distinguish signals induced by axions or GWs, from features of the noise spectra measured in the experiment.

\section{Conclusions}
\label{sec:conclusions}

In this work, we studied the evolution of light polarization as it propagates through the background of axion DM or that of a passing plane GW. These effects are closely related to the ordinary Faraday effect, according to which the polarization vector of linearly polarized light rotates around the direction of propagation, with the rotation angle being proportional to the strength of the magnetic field. In the presence of an axion field, a comparable effect arises, with the rotation angle instead being proportional to the axion abundance and the axion-photon coupling. Likewise, when light propagates through the background of a passing GW, the polarization changes, though the effect goes beyond a simple rotation about the direction of motion.  Within the framework of geometric optics, we provided a unified treatment of these effects, demonstrating that the polarization evolution in all cases originates from the same underlying physics. This approach provides a synergy between searches for axion DM and those for GWs.
 
Following this framework, we demonstrated that these effects can be exploited in the optical cavities of the ALPS II experiment, initially designed to observe light-shining-through-the-wall induced by axions, but easily adaptable to measure polarization effects from other sources. Specifically, we showed that by measuring the change in polarization of its laser, axion DM with masses in the range $m \sim 10^{-9} - 10^{-7} \, \mathrm{eV}$ can be probed. This search is competitive with other laboratory-based experiments and with astrophysical searches, particularly near resonance frequencies. Moreover, we discussed how this method could achieve a more broadband sensitivity by introducing a QWP into the setup, which allows the polarization effects to be resonantly enhanced by the cavity at frequencies where they otherwise would not be.  The combined results for axion DM are summarized in Fig.~\ref{fig:ax_sens2}. 

A natural application of this method is searching for GWs in optical cavities. In this work we showed that the ALPS II experiment can probe  strain sensitivities of around $10^{-14} \, \mathrm{Hz}^{-1/2}$ in the frequency range of 20~MHz to 100~GHz.  In this way, with only minor modifications, the ALPS II experiment may be able to explore currently unconstrained parameter space by other strategies proposed to search for high-frequency GWs. In the future, other type of experiments, currently in R\&D phase, could significantly enhance these prospects, though such advancements may not occur in the near term.
We also demonstrated how to optimize this GW search strategy. As in the axion case, introducing a QWP significantly improves sensitivity, especially if the GW itself is polarized. Without a QWP, the signal from the $h_+$ polarization component is suppressed. This behavior is reminiscent of similar phenomena predicted in other experiments searching for high-frequency GWs in axion detectors. 

The combined results for GWs are summarized in Fig.~\ref{fig:h0_gw}. The advantage of interferometric sensing schemes that measure the relative length changes induced by GWs acting on two orthogonally oriented arms is clear in this plot, even for those not using arm cavities. This can be intuitively understood as the interferometric sensing schemes having an extra factor of $\ell/\lambda$ in the cavity response function due to the GW acting on the long `lever arm' of the cavity in comparison to polarimetric sensing schemes. Nevertheless, the simplicity of the experimental design proposed here could enable its rapid implementation on an existing optical cavity such as those used in ALPS\,II. This makes this one of the most promising concepts to probe the region of frequency space above 100\,MHz in the near-term future. 
\section*{Acknowledgments}

We thank Valerie Domcke, Aldo Ejlli, Pedro Schwaller, and Benno Willke  for useful discussions.  C.G.C. is supported by a Ramón y Cajal contract with Ref.~RYC2020-029248-I, the Spanish National Grant PID2022-137268NA-C55 and Generalitat Valenciana through the grant CIPROM/22/69.
L.M is supported by the the Spanish National Grant PID2022-137268NA-C55 and by the Generalitat Valenciana through the grant CIACIF/2023/91. 
The work of A.D. Spector and A.R. is partially funded by the Deutsche Forschungsgemeinschaft
(DFG, German Research Foundation) under Germany’s Excellence Strategy - EXC2121Quantum
Universe - 390833306 and under - 491245950. This article/publication is based upon work from
COST Action COSMIC WISPers CA21106, supported by COST (European Cooperation in Science
and Technology).
\appendix

\section{Evolution of light polarization in the geometric-optics limit}
 \label{app:A} 
Maxwell's equations in flat spacetime can be written as 
\begin{align}
\partial_\nu F^{\mu\nu} =  J^\mu\,, && \partial^\lambda F^{\mu\nu}+\partial^\mu F^{\nu\lambda}+\partial^\nu F^{\lambda\mu}=0  \,.
\label{eq:appMaxwell}
\end{align}
As is well-known, both equations lead to the wave equation
\begin{align}
\partial^2 F^{\mu\nu} =  -\partial^\mu J^\nu+\partial^\nu J^\mu \,.\end{align}
When the wavelength of electromagnetic fluctuations, $\lambda$, is much smaller than the characteristic length scales, $d$, associated with the background over which these fluctuations propagate, the equations can be solved using the geometric limit, also known as the Eikonal approximation. In practice, as shown in the following, this is equivalent to solve for the momentum and polarization of photons as they propagate in the background.  The limit of small wavelength can be formally obtained by writing the field and the current as %
\begin{align}
F^{\mu\nu} \equiv (f^{\mu\nu}+f_1^{\mu\nu} \epsilon+f_2^{\mu\nu} \epsilon^2+\ldots) e^{i \theta/\epsilon }  \,,&&
J^\mu \equiv (j^\mu+j_1^\mu \epsilon+j_2^\mu \epsilon^2+\ldots) e^{i \theta/\epsilon }  \,.
\end{align}
We then solve for Maxwell's equations expanding on $\epsilon$. As explained in the text, $\epsilon$ is a fictitious parameter such that a term multiplied by $\epsilon^n$ in the expansion is of order $(\lambda/d)^n$. The leading term in this expansion is the geometric-optics limit of electromagnetism~\cite{Maggiore:2007ulw}. With this in mind, let us note that 
\begin{align}
\partial_\nu F^{\mu\nu} =  i \left( f^{\mu\nu} \partial_\nu \theta \right)\frac{1}{\epsilon}+{\cal O}(\epsilon^0)\,,&& 
\partial^\mu J^\nu =  i \left( j^\nu \partial^\mu \theta \right)\frac{1}{\epsilon}+{\cal O}(\epsilon^0) \,,
\label{eq:appwaveEq}
\end{align}
as well as 
\begin{eqnarray}
\partial_\rho \partial^\rho F^{\mu\nu} &=& \left(- f^{\mu\nu} \partial_\rho \theta \partial^\rho \theta\right) \frac{1}{\epsilon^2}+\left(- f_1^{\mu\nu} \partial_\rho \theta \partial^\rho \theta+2i \partial^\rho f^{\mu\nu} \partial_\rho \theta + i f^{\mu\nu} \partial_{\rho}\partial^{\rho}\theta  \right) \frac{1}{\epsilon} +{\cal O}(\epsilon^0)\nonumber\,.
\end{eqnarray}
Introducing $k_\mu  \equiv \partial_\mu \theta$, Eq~\eqref{eq:appMaxwell} imply that 

\begin{align}
k_\mu j^\mu = 0\,, && 
k_\nu f^{\mu\nu} = 0 \,, && \text{and}&&
k^\lambda f^{\mu\nu}+k^\mu f^{\nu\lambda}+k^\nu f^{\lambda\mu}=0\,,
\label{eq:appgeometric1}
\end{align}
while the wave equation gives 

\begin{align}
k_\mu k^\mu = 0\,, &&\text{and} &&
k^\rho \partial_\rho f^{\mu\nu} = -\frac{1}{2} f^{\mu\nu} \partial^\rho k_\rho- \frac{1}{2} k^\mu j^\nu+\frac{1}{2} k^\nu j^\mu\,.
\label{eq:appgeometric2}
\end{align}
The vector $k^\mu$ can thus be interpreted as  four-momentum of photons while $E^i\equiv f^{0i}$ determine their polarization.  Note in particular that the last relation in Eqs.~\eqref{eq:appgeometric1}  gives the other components in terms of $E^i$ as $ f^{ij}=( k^i E^{j} -k^j E^{i})/k^0$.  Similarly, from Eq.~\eqref{eq:appgeometric2}, it follows that%
\footnote{
In more detail, Eq.~\eqref{eq:appgeometric2} for the electric field reads
\begin{align}
k^\rho \partial_\rho {\bf E}= -\frac{1}{2}\, {\bf E}  \, \partial^\rho k_\rho- \frac{1}{2} k^0 {\bf j}+\frac{1}{2} {\bf k} j^0\,,&&
\text{and}&&
k^\rho \partial_\rho ({\bf E}^*\cdot {\bf E})= -\, {\bf E}^*\cdot {\bf E}  \, \partial^\rho k_\rho-  k^0 {\rm Re}\left(  {\bf E}^* \cdot {\bf j} \right)\,.
\end{align}
%
%
%
%
%
%
%
Together they lead to
\begin{eqnarray}
k^\rho \partial_\rho {\bf e}= - \frac{1}{2|{\bf E}|} k^0 {\bf j}+\frac{1}{2|{\bf E}|} {\bf k} \,j^0+ \frac{1}{2 |{\bf E}|} {\rm Re}\left(  {\bf e}^* \cdot {\bf j} \right)k^0 {\bf e}\,,
\end{eqnarray}
which gives rise to Eq.~\eqref{eq:appflatfinal0} after accounting for $k_\mu j^\mu=0$ and $k_\mu k^\mu=0$.
} 
the unit vector, $ {\bf e}={\bf E}/ |{\bf E}|$, changes according to

\begin{align}
 k^\rho \partial_\rho {\bf e} =
 - \frac{1}{2|{\bf E}|} k^0 \left({\bf j}-(\hat{\bf k}\cdot {\bf j})\hat{\bf k}\right)+ \frac{1}{2 |{\bf E}|} {\rm Re}\left(  {\bf e}^* \cdot {\bf j} \right)k^0 {\bf e}\,.
 \label{eq:appflatfinal0}
\end{align}
Motivated by the cases discussed below, we assume ${\rm Re}\left(  {\bf e}^*\cdot{\bf j} \right)=0$ from now on and therefore omit the last term. Photon trajectories are curves, $x^\mu(\ell)$, with $k^\mu$ as their tangent vector. Therefore 
\begin{align}
k^\mu = \frac{dx^\mu}{d\ell} \,, && \text{and}&& \frac{dx^\mu}{dt} = \frac{k^\mu}{k^0}\,.
\end{align}
Using this, Eq.~\eqref{eq:appflatfinal0} gives the evolution of the polarization vector as  
\begin{align}
\frac{d{\bf e}}{dt} = 
 - \frac{1}{2|{\bf E}|} \left[{\bf j}-(\hat{\bf k}\cdot {\bf j})\hat{\bf k}\right]\,. 
  \label{eq:appflatfinal}
\end{align}

Let us apply this to axion birefringence. Electromagnetic waves propagating in slowly-changing axion background $a(t)$ induce an effective current given
\begin{eqnarray}
{\bf J}=g_{a\gamma\gamma} \dot{a}(t) \mathbf{B}\,.
\label{eq:axionsj}
\end{eqnarray}
This follows from e.g.~the Lagrangian in Eq.~\eqref{eq:Maxwell}. Here, ${\bf B} = \hat{\bf k} \times {\bf E} $ is the magnetic field associated with the electromagnetic wave, and we use again the notation so that $E^i \propto f^{0i}$. Although this is not an actual current of charged particles, the effect of the axions is mathematically equivalent to that of an ordinary current given by Eq.~\eqref{eq:axionsj}.  Note that ${\rm Re}\left(  {\bf e}^*\cdot{\bf j} \right)$ vanishes since ${\bf E}^*\times{\bf E}$ is always purely imaginary, and that  $\hat{\bf k}\cdot {\bf j}=0$. As a result, Eq.~\eqref{eq:appflatfinal0} predicts 
\begin{align}
\frac{d{\bf e}}{dt} = 
 - \frac{1}{2}g_{a\gamma\gamma}  \dot{a}(t) \, \hat{\bf k} \times {\bf e}\,.
 \label{eq:axions}
\end{align}

For linear polarizations, this equation states that the vector ${\bf e}$ rotates with ``angular velocity'' $g_{a\gamma\gamma}  \dot{a}(t)/2$ around the direction $\hat{\bf k}$, while its absolute magnitude and the angle which it makes with this direction remain fixed. For instance, suppose an initial condition such that $ {\bf e} =(1,0,0)$ with a photon propagating in the $z$ direction. After some distance the polarization vector rotates by an angle of
\begin{equation}
\beta = - \frac{1}{2}g_{a\gamma\gamma}  \int dt  \, \dot{a}(t) \,.
\label{eq:appRotationAngleAxion}
\end{equation}

On the other hand, for right (left) circular polarizations
$
\hat{\bf k} \times {\bf e} = -i \lambda \, {\bf e}
$
with $\lambda$ is $+1$ ($-1$). In this case, Eq.~\eqref{eq:axions} predicts an evolution given by a phase dependent on the circular polarization.  This is the origin of the term birefringence. Clearly, all this resembles the Faraday effect, a magneto-optical phenomenon where the polarization plane of linearly polarized light rotates as it propagates through a material in the presence of a magnetic field.

\section{Geometric-optics in curved spacetimes}
\label{app:B}

We now extend the previous results to curved spacetimes, and in particular, to GWs. As in the case of axions, the effect of a GW in the presence of an electromagnetic field can be effectively described by an effective current in Minkowski spacetime, see e.g.~\cite{Domcke:2022rgu}. Although the calculation using that formalism is straightforward, here we instead take $J^\mu=0$ and note the derivation of Eqs.~\eqref{eq:appgeometric1} and \eqref{eq:appgeometric2} was done without assuming $\partial^\alpha \partial^\beta =\partial^\beta \partial^\alpha$ and are therefore equally valid for covariant derivatives. In an arbitrary spacetime we then have  

\begin{align}
k_\mu k^\mu = 0\,, &&
k_\nu f^{\mu\nu} = 0 \,, && 
k^\lambda f^{\mu\nu}+k^\mu f^{\nu\lambda}+k^\nu f^{\lambda\mu}=0\,,
&&
k^\rho \nabla_\rho f^{\mu\nu} = -\frac{1}{2} f^{\mu\nu} \nabla^\rho k_\rho\,.
\label{eq:appgeometric3APP}
\end{align}

Following an approach similar to the one above, we will use these expressions to  write an equation for the evolution of $e^{i} =f^{0i}/|f|$ with $|f| = \sqrt{ f^{0i}f^*_{0i}}$. Concretely, 
for $f^{0i}$  Eq.~\eqref{eq:appgeometric3APP} gives
\begin{eqnarray}
k^\rho (\partial_\rho f^{0i}+\Gamma^{0}_{\rho \lambda}f^{\lambda i}+\Gamma^{i}_{\rho \lambda}f^{0\lambda}) = -\frac{1}{2} f^{0i} \nabla^\rho k_\rho\,.
\label{eq:intermediate}
\end{eqnarray}
Using
$k^0 f^{\lambda i}=k^\lambda f^{0 i} -k^i f^{0 \lambda}$, we obtain
\begin{eqnarray}
k^\rho \left(f^*_{0i}\partial_\rho f^{0i}+ \frac{1}{k^0}\Gamma^{0}_{\rho \lambda}k^\lambda f^*_{0i} f^{0 i} -0 
+f^*_{0i}\Gamma^{i}_{\rho \lambda}f^{0\lambda}\right) = -\frac{1}{2} f^*_{0i}f^{0i} \nabla^\rho k_\rho\,.
\end{eqnarray}
Similarly, for the covariant components 
\begin{eqnarray}
k^\rho \left(f^{0i}\partial_\rho f^*_{0i}- \frac{1}{k_0}\Gamma^{\lambda}_{\rho 0}  k_\lambda f^{0i}f^*_{0i}+0
-f^{0i}\Gamma^{ \lambda}_{\rho i}f^*_{0\lambda}\right) = -\frac{1}{2} f^{0i}f^*_{0i} \nabla^\rho k_\rho\,.
\end{eqnarray}
Adding both, it is found that
\begin{eqnarray}
\!\!\!\!\!\!\!\!\!\!\!\!
k^\rho \left(\partial_\rho |f|+ \frac{1}{2k^0}(\Gamma^{0}_{\rho \lambda}k^\lambda+\Gamma^{\lambda}_{\rho 0}  k_\lambda ) |f|+\frac{1}{2|f|}f^*_{0i}\Gamma^{i}_{\rho \lambda}f^{0\lambda}
-\frac{1}{2|f|}f^{0i}\Gamma^{ \lambda}_{\rho i}f^*_{0\lambda}\right) = -\frac{|f|}{2} \nabla^\rho k_\rho\,.
\end{eqnarray}
The last two terms on the left hand side cancel each other, while the first two are equal to each other. 
Hence
\begin{eqnarray}
k^\rho \left(\partial_\rho |f|+ \frac{1}{k^0}\Gamma^{0}_{\rho \lambda}k^\lambda  |f| \right)= -\frac{|f|}{2} \nabla^\rho k_\rho\,.
\end{eqnarray}
Combining this with Eq.~\eqref{eq:intermediate} we obtain
\begin{eqnarray}
k^\rho \left(\partial_\rho e^{i} -\frac{1}{k^0} \Gamma^{0}_{\rho \lambda}k^i e^{\lambda}+\Gamma^{i}_{\rho \lambda}e^{\lambda}\right) = 0\,.
\end{eqnarray}
As explained in the main text, this gives rise to Eq.~\eqref{eq:finalGW} along null geodesics.
\bibliographystyle{JHEP}
\bibliography{main}

\end{document}